\documentclass[aps,pra,twocolumn,superscriptaddress,nofootinbibt,longbibliography]{revtex4-1}
\usepackage[main=english,german,ngerman]{babel}
\usepackage[T1]{fontenc}
\usepackage{hyphenat}
\usepackage{hyperref}
\usepackage{amsmath,amssymb}
\usepackage{amsfonts}
\usepackage{epsfig,pstricks,graphicx}
\usepackage{bm}
\usepackage{physics}
\usepackage{color}

\definecolor{armygreen}{rgb}{0.29, 0.33, 0.13}
\usepackage{color}
\usepackage{array}
\def\id{{\rm 1\kern-.22em l}}

\newcommand{\av}[1]{\left \langle #1 \right\rangle}

\usepackage[normalem]{ulem} 
\usepackage{soul} 
\usepackage{epigraph}

\newcommand{\cor}[1]{{\color{black} #1}}

\newcommand{\orcidicon}[1]{\href{https://orcid.org/#1}{\includegraphics[height=\fontcharht\font`\B]{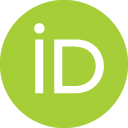}}}

\begin{document}

\title{Finite-size criticality in fully connected spin models on superconducting quantum hardware}

\author{Michele~Grossi\orcidicon{0000-0003-1718-1314}}
\email{michele.grossi@cern.ch} 
\affiliation{European Organization for Nuclear Research (CERN), Geneva 1211, Switzerland}

\author{Oriel~Kiss\orcidicon{0000-0001-7461-3342}}
\affiliation{European Organization for Nuclear Research (CERN), Geneva 1211, Switzerland} 

\affiliation{Department of Particle and Nuclear Physics, University of Geneva, 1211 Geneva, Switzerland}

\author{Francesco De Luca}
\affiliation{Department of Physics, University of Trieste,  34127 Trieste, Italy}

\author{Carlo Zollo}
\affiliation{Department of Physics, University of Trieste,  34127 Trieste, Italy}

\author{Ian Gremese}
\affiliation{Department of Physics, University of Trieste,  34127 Trieste, Italy}

\author{Antonio~Mandarino\orcidicon{0000-0003-3745-5204}}
\email{antonio.mandarino@ug.edu.pl}
\affiliation{International Centre for Theory of Quantum Technologies, University of Gdańsk, Jana Bażyńskiego 1A,
80-309 Gdańsk, Poland}

\begin{abstract}
The emergence of a collective behavior in a many-body system is responsible of 
the quantum criticality separating different phases of matter. 
Interacting spin systems in a magnetic field offer a tantalizing opportunity to 
test different approaches to study quantum phase transitions. 
In this work, we exploit the new resources offered by quantum algorithms to detect 
the quantum critical behaviour of fully connected spin$-1/2$ models. 
We define a suitable Hamiltonian depending on an internal anisotropy parameter $\gamma,$ that  allows us to examine three paradigmatic examples of spin models, whose lattice is a fully connected graph. 
We propose a method based on variational algorithms run on superconducting transmon qubits to detect the critical behavior 
for systems of finite size. We evaluate the energy gap between the first excited state and the
ground state,  
the magnetization along the easy-axis of the system, and the spin-spin correlations.
We finally report a discussion about the feasibility of scaling such approach on a 
real quantum device for a system having a dimension such that classical simulations 
start requiring significant resources.

\end{abstract}

\maketitle

\section{Introduction}
The abrupt change of the system properties during a phase transition 
has always paved the way to the advancement of our 
understanding of nature in both fundamental and applied aspects. 
The phase transition mechanism, in the limit of an infinite number of particle composing the system, 
has been successfully addressed within the formalism of the renormalization group \cite{zinn2007phase, pelissetto2002critical}. 
Quantum phase transitions are the cornerstone of a great variety of 
groundbreaking theories ranging from the Higgs mechanism for mass generation in 
high\hyp energy physics \cite{higgsmechanism2,higgsmechanism}, to the superfluid and superconducting phase of matters in low\hyp energy physics \cite{BCS,tilley2019superfluidity}, 
and nowadays their exploitation is getting attention also in the context of quantum technologies \cite{manybody_rev, manybody_rev2}. 

Given a Hamiltonian $H(\vec \lambda)$, describing a system constituted by $N$ interacting particle, 
it exhibits a continuous (or second order) quantum phase transition, whether in the limit $N \rightarrow \infty,$ the 
gap between the ground state energy and that of the first excited state vanishes for a certain value of the internal parameters $\vec \lambda$. 
This value corresponds to the critical point of the model and, in contrast to any classical model, it can also exist for zero temperature \cite{sachdev_2011, RevQPT}. 
Nevertheless, assuming a diverging number of particle is well motivated and substantiated. 
The relaxation of such assumption prompts the study of finite-size corrections to such transition \cite{finiteCorr1,finiteCorr2} that can show unprecedented results \cite{finitesize1, finitesize2, finitesize3}. 
With abuse of notation we write that a quantum phase transition occurs in a system with finite $N$, 
whenever for a value of $\vec \lambda$, a crossing between the energies of the ground state and the first excited level is observed.  
This is in contrast to what one would expect by a semi-classical approach in which the finite 
size is responsible to suppress the symmetry breaking mechanism associated with the second-order phase transition \cite{finiteCorr1,finiteCorr3}. 
Criticality of quantum system requires an exponential number of degrees of freedom that makes the problem quickly intractable. The advancement of machine learning techniques has been of paramount importance for the determination of macroscopic phases of matter and efficient quantum state representation. 

With the advent of quantum techniques in machine learning, phase diagram of different systems have been obtained, such as a cluster Ising or the Bose\hyp Hubbard model at zero temperature. The former uses a supervised learning approach where the states are classified according to classical labels using a quantum convolutional neural network \cite{Cong_19, Wallraff_22_QPD}  while the latter discovers the phases in an unsupervised way using anomaly detection \cite{PRR_anomaly_detection}. 
The intersection between machine learning and quantum techniques applied to physical systems is rapidly increasing,  not only obtaining information about critical point of a system is pursued but also general dynamical simulations are important testbeds. In \cite{caroQML} the authors rigorously analyze the requirements of an algorithm in terms of training data and define generalization bounds for their effective execution on current quantum device. For an overview of the state of the art and future perspectives for quantum simulation, looking at possible quantum advantage in specific applications we refer to \cite{zollerqadv}.

We consider the Lipkin-Meshkov-Glick (LMG) model, a fermionic model that served as testbed for many-body approximations in different 
fields \cite{LMG1,LMG2,LMG3}. 
Due to the possibility of mapping this model into a $N$ spin$-1/2$ system, we will study its criticality 
with current quantum computation techniques on real Noisy-Intermediate-Scale Quantum (NISQ) devices \cite{Preskill2018quantumcomputingin,RMP_NISQ}.

Here, we propose a way to study the finite-size critical behavior of the system with the freshly introduced methods of quantum machine learning. Our motivation is threefold. Firstly, very few examples of such methods have been used to study magnetic systems \cite{VQE_QPT}. Secondly, the finite size critically with regard to this model is of interest to the molecular magnetism community \cite{LMG_nano}. Lastly, and a bit ambitiously, we would like to pave the way to a feasible application of the  nowadays-available NISQ hardware as a tool to simulate physics with (quantum) computers. In fact, before addressing more complex magnetic systems, we have chosen as a testbed a magnetic model that due to its symmetries served 
to validate geometrical methods developed in the framework of 
quantum information \cite{LMG_susc1, LMG_susc2, LMG_geom} as signature of quantum phase transitions.  However, the class of models that can be studied are not so trivially integrable to be considered as a mere academic exercise.

A simplified version of the spin model that we consider 
has been tackled in recent papers \cite{LMG_QC1,LMG_QC2,LMG_QC3,LMG_QC4} for system up to four spins and the analysis is carried on noiseless simulators.
This allows us to contribute to the validation of a \emph{heuristic} approach such as the Variational Quantum Eigensolver (VQE) \cite{vqe_original, Rev_VQE}
in the challenging research field of the statistical physics of finite-size models on a lattice.  

The remainder of this manuscript is structured as follows: in Sec.\ref{sec:methods} we provide a short but comprehensive introduction to the VQE technique, then we focus on the definition of the wavefunction ansatz in terms of design and trainability \cor{of quantum circuits} and we provide an overview of the adopted error mitigation techniques, with ad-hoc consideration for the specific Hamiltonian. \cor{After introduucing all the tools, we} terminate the section with the definition of the LMG Hamiltonian of our critical system.
In Sec~\ref{sec:results}, we substantiate our approach showing simulated results obtained under ideal condition with quantum simulator as well as evidences collected on real quantum device.
In Sec.~\ref{sec:excited}, we provide a numerical interpretation and analytical derivation of higher order excited states for the LMG model, as well as their realisation with the variational algorithm.
Finally in Sec.~\ref{sec:discussion}, we summarise the outcomes of this work, discussing the quality of the results with an estimation about the actual feasibility of studying proposed models-like on NISQ devices.

\section{Methods}
\label{sec:methods}
In this section, \cor{before introducing} the Hamiltonian of a LMG critical system and its behaviour in the thermodynamic limit, we review the quantum computational techniques that we employ to assess the critical behavior of the system. 

\subsection{Variational Quantum Eigensolver}
\label{subsec:VQE}
The VQE, proposed by \textcite{vqe_original}, is a variational quantum algorithm \cite{variational_quantum_algorithm} used to find the ground state of a Hamiltonian $H$ by using the Rayleigh\hyp Ritz variational principle. 
\cor{This variational method has been widely applied in quantum chemistry \cite{UCC-chemistry,VQE_Gambetta,magnetogrossi,Panos_excitation_preserving}, nuclear physics \cite{Papenbrock-Deuterium,Be8-VQE,VQE_Li6} and in spin systems \cite{VQE_magnet,VQE_ising, LMG_QC1,LMG_QC2,LMG_QC3,LMG_QC4,Hlatshwayo:2022yqt}.}

Concretely, a parametrized  wavefunction $\ket{\psi(\theta)}$ \cite{PQC} is prepared on a quantum computer and its parameters updated to minimize the energy 
\begin{equation}
    E_0 \leq 
    \frac{    \bra{\psi(\theta)}H\ket{\psi(\theta)}}{\braket{\psi(\theta)}}
\end{equation}
where the normalization factor at the denominator can be dropped if the wavefunction is normalized.

The design of the wavefunction ansatz is of importance for the trainability and accuracy of the results and is an active area of research. Some systems, typically written in the second quantisation formalism, allow physically motivated ans\"atze, for instance based on unitary coupled cluster \cite{UCC-chemistry,Be8-VQE,VQE_Li6,anand2022quantum,lee2018generalized}.
\cor{In this setting the related} quantum circuits are usually deep, require an increased connectivity, and are therefore difficult to implement on near term quantum devices. On the other hand, hardware efficient ansatz (HEA) \cite{VQE_Gambetta} are tailored to the device and are consequently shallow enough to minimize the effects of noise and decoherence. 
 Despite working with shallow circuit, in general HEA may suffer from scalability issues due to the increasing number of parameters to optimise, leading to untrainability issues, namely barren plateau \cite{Barren_platea_McClean}.
 An alternative direction to optimise the choice of the ans\"atze is the possibility of exploiting symmetries in the system. As recently proposed in \cite{equivar} it is possible to work with equivariant ans\"atze which might mitigate the barren plateau problem. However, as underlined by the authors, there exist always a trade-off between the equivariance and expressivity of the parametric circuit.

More recently, the ADAPT-VQE \cite{ADAPT-VQE}, which builds the ansatz by iteratively adding a term from an operator pool bringing the best improvement, has been proposed as a way to build optimal circuits. Even if the picking action can be implemented in a parallel fashion, it can be expensive for current devices,  time and resources\hyp wise. Consequently, we will focus on fixed hardware efficient ans\"atze, which are constructed with single qubit  rotations around the y\hyp axis, and CNOT interactions with linear connectivity.

The VQE can be extended to compute excited states as well. The method adopted here is the one proposed by \textcite{Excited-states}, called \emph{variational quantum deflation} (VQD), which first computes the ground state and then looks for the state minimizing the energy while being orthogonal to the, previously determined, ground state. This procedure can be generalize for the $k$-th excited state in an iterative fashion. In practice, the following loss function is minimized:
\begin{equation}
\label{eq:VQD}
F(\theta_k) = \bra{\psi(\theta_k)}H\ket{\psi(\theta_k)} + \sum_{i=0}^{k-1}\beta_i \bra{\psi(\theta_k)}\ket{\psi(\theta_i)},
\end{equation}
where we assume, for simplicity, that the states are normalized. The wavefunction $\ket{\psi(\theta_i)}$ corresponds to the $i$-th excited state, and $\beta_i$ hyperparameters to be tuned. It has been shown \cite{Excited-states}, that $\beta_i$ has to be greater than the energy gap between the states $i$ and $i+1$ to ensure that the wavefunction converges to the correct excited state. Additional techniques, based on the quantum equation of motion \cite{Ollitrault_excited_state}, using a discriminator \cite{DVQE} or constraining the ansatz around the state of interest \cite{SSVQE_Mitarai} have been proposed in the literature but will not be considered here.
\subsection{Ansatz}
\label{sec:ansatz}
\begin{figure}
    \includegraphics[scale=0.6]{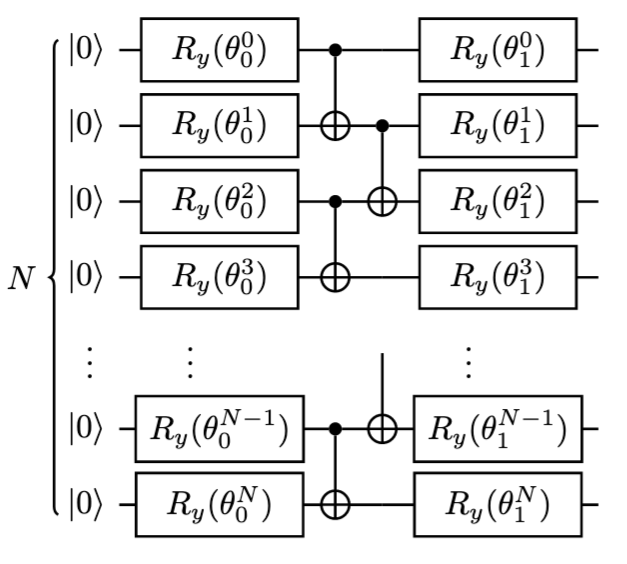}
    \caption{\cor{Representation of the chosen ansatz to design the parametric function $\psi(\theta)$, namely the Hardware efficient ansatz, composed of $R_y(\theta)$ rotations and CNOTs between neighboring qubits executed in parallel. We sketch here a single layer of the circuit, corresponding to $D=1,$ higher values of $D$ correspond to sequential repetitions of such circuit.} }
    \label{fig:ansatz}
\end{figure}

We use a simple hardware efficient ansatz \cite{VQE_Gambetta}, which can be run on NISQ devices without an overhead due to circuit transpilation.  For instance, we use $D$ repetitions of a layer consisting of free rotations around the $y$\hyp axis $R_y(\theta)=e^{-i\theta \sigma_y/2}$, where $\sigma_y$ is the $y$ Pauli matrix, CNOT gates with linear connectivity and a final rotation layer before the measurements. Since the depth of the circuits grows as $\mathcal{O}(N)$ due to the linear entanglement, this ansatz fails on hardware when performing error mitigation based on noise scaling. We therefore adapt the ansatz to grow as $\mathcal{O}(1)$, by applying the CNOT gates in parallel, on the two following groups of qubits 
\begin{align}
    \nonumber 
    \{(i,i+1)& \text{ for \emph{i} even} \}, \\
    \nonumber
     \{(i,i+1)& \text{ for \emph{i} odd} \},
\end{align}
 thus considerably reducing the depth of the circuits. We observe a small decrease in the accuracy on the simulator compared to the linear entanglement scheme, but an increase on the hardware due to the depth's reduction. We choose the minimal case $D=1$, as depicted in Figure \ref{fig:ansatz}, when running on quantum hardware, while pushing for maximal performance on the simulator by allowing larger $D$.
\subsection{Error Mitigation}
\label{sec:mitigation}
Error mitigation methods are used to diminish the effect of the hardware noise on the results. Unlike error correction, these strategies are used in the post\hyp processing steps on the raw data. Two complementary techniques, measurement error mitigation (MEM) and zero noise extrapolation (ZNE), are used to mitigate the readout and two\hyp qubit gate errors, respectively. 

For MEM, we follow \textcite{MEM} and individually invert the error matrices 
\begin{equation}
M^k=
\begin{pmatrix}
P^{(k)}_{0,0} &P^{(k)}_{0,1} \\P^{(k)}_{1,0} &P^{(k)}_{1,1}
\end{pmatrix}
\end{equation} 
 and used them to calibrate the samples. Here, $P^{(k)}_{i,j}$ is the probability of the $k$-th qubit to be in state $j\in\{0,1\}$ while measured in state $i\in\{0,1\}$. The probabilities of measuring 0 or 1
\begin{equation}
    \vec{S}^k = \begin{pmatrix}
    P_0 \\P_1
    \end{pmatrix}
\end{equation} obtained by measuring the $k$-th qubit are corrected as follow
\begin{equation}
    \vec{S}^k_{\text{corrected}} = (M^k)^{-1}\vec{S}^k.
\end{equation}
 While this only corrects uncorrelated readout errors, it is argued in Ref.~\cite{MEM} that they are the predominant ones. Moreover, this strategy can be scaled for arbitrary number of qubits and only has a $\mathcal{O}(1)$ overhead in the number of circuit execution.
 In the ZNE \cite{ZNE,error-mitigation} scheme, the CNOT noise is artificially stretched and the results are then extrapolated to the noiseless regime. More precisely, the energy is estimated multiples time for different scaling factor $k\in \{1,2\}$, and then a fit is performed to extrapolate up to the $k=0$ value. In practice, the noise is stretched by replacing every CNOT in the circuit, by $2k-1$ CNOT gates. The $2k-2$ additional CNOTs cancel each other, leaving the circuit unchanged. However, by adding barriers between them, preventing the CNOTs to be cancelled in the transpilation phase, the noise is artificially stretched. \textcite{Richardson} originally used a linear fit for the extrapolation, however, the considerable effect of the noise in NISQ devices increases the risk of overshooting. Consequently, an exponential fit $f(x) = ae^{bx}$ is instead used, where $a$ and $b\in \mathbb{R}$ are fitted to the energies $E$ using least\hyp square regression. To improve the results, $E$ is scaled before the fit
 \begin{equation}
     E\mapsto \frac{E-s}{s}
 \end{equation}
 and scaled back afterwards, with $s$ being an estimate of the exact energy. 
In ZNE the scaling of sampling required could also be exponential, at least in some cases \cite{Takagi_2022}. Of course this is just a theoretical upper bound in measurement but in practise we use a scalable MEM method which requires only 2 circuits as a sufficient quantity to mitigate the noise.
 
 It is important to make sure that the total runtime of the noise\hyp scaled circuits does not exceed the coherence time of the device, which would destroy any useful information. For instance, we only considered $k=1$ and $k=2$, since for higher $k$ the results were not reliable anymore. Also the ansatz definition plays an important role, as described in sec.\ref{sec:ansatz}. With the construction of Figure \ref{fig:ansatz}, the CNOT gates can be run in parallel, thus shortening the runtime significantly. This can be done using additional qubits available on the device to run all the noise\hyp scaled circuits in parallel, reducing the total number of circuit execution, however for sake of meticulousness one can note that in principle this strategy might result in additional cross-talks, even if, looking at IBM current hexagon topology, one can really minimize this effect.
 
\subsection{The Lipkin-Meshkov-Glick Model}
\label{sec:LMG}
The LMG model was introduced in \cite{LMG1,LMG2,LMG3} to describe a system of $N$ fermions, 
whose state space is made of two degenerate shells with two fixed energy levels. 
Each shell has degeneracy $N$ and can accommodate all of the $N$ particles, thus resulting in a total of a $2^N$-dimensional state space. 
Due to the symmetry of the Hamiltonian with the total spin, the low energy states are in the subspace of maximum spin, which has a dimension that scales linearly with $N$. 
However, to substantiate our approach we do not consider only the maximum spin sector of the Hamiltonian but we use the VQE on the full space as outlined in Section \ref{subsec:VQE}. 
Via a Jordan-Wigner transformation \cite{lieb1961}, the LMG model can be mapped into a system of interacting spins. Moreover, 
in the thermodynamic limit $N\rightarrow \infty$, it is solvable via a two-boson Holstein Primakoff transformation \cite{LMG_MB1}. 
This peculiarity made it one of the most used model to understand many problems of interest in physics, from nuclear to condensed-matter physics. 

Considering that we will study our model on qubit-based quantum computers, it is natural and convenient to use the following 
expression for the LMG Hamiltonian: 
\begin{equation}
\label{eq:LMGhamiltonian}
H= -\frac{1}{N}\sum_{i<j}^N \sigma^i_x \sigma^j_x + \gamma \sigma^i_y \sigma^j_y - B \sum_{i=1}^N \sigma^i_z. 
\end{equation}
This Hamiltonian describes a system of $N$ spins in a fully connected planar graph, immersed in a transverse magnetic field $B.$
The first sum in Eq.\eqref{eq:LMGhamiltonian} accounts for an anisotropic interaction in the $x-y$ plane that couples each spin with all the other ones with the same strength, 
an archetype and exemplary version of any long-range interaction. 
Different coupling strengths along the two planar direction are taken into account via the anisotropy parameter $0 \leq \gamma \leq 1.$
From physical perspective, this type of Hamiltonian has been implemented on various 
platforms \cite{LMG_impl1,LMG_impl2,LMG_impl3,LMG_impl4} to design feasible quantum technologies applications \cite{LMG_QT1,LMG_QT2,LMG_QT3,LMG_QT6,LMG_QT7,LMG_QT8}.  

The system is known to be critical and shows, in the thermodynamic limit, a second order phase transition 
between a broken-symmetry (disordered) phase for $B<1$ and an ordered phase for $B \geq 1$, with a critical value of 
the external magnetic field $B=B_c= 1.$
Usually, the Lipkin model is used to denote a limiting and easily diagonalizable case of the LMG model \cite{LMG4}.
Introducing the set of collective-spin operators  
$S_\alpha = \frac12 \sum_{i=1}^N \sigma^i_\alpha $, and setting $\gamma= 1$ in Eq.\eqref{eq:LMGhamiltonian}, we have:
\begin{equation}\label{eq:LMGDickeHamiltonian}
H= -\frac{2}{N} (\mathbf{S}^2 - S_z^2)- 2 B S_z. 
\end{equation}
The Hamiltonian in Eq.\eqref{eq:LMGDickeHamiltonian} is diagonal in the Dicke basis  $ \ket{j, m}$ formed by the simultaneous eigenvectors
of $\mathbf{S}^2 \ket{j, m}=  j(j+1)\ket{j, m} $ and  $ S_z \ket{j, m}= m \ket{j, m}$.
Due to the fact that the interaction term commutes with the free energy term, the Lipkin model with $\gamma=1$ 
belongs to a different universality class of the general model described by the Hamiltonian in Eq.\eqref{eq:LMGDickeHamiltonian}, see \cite{LMG_MB1}. 
In particular, it has been shown to belong to the same class of the superradiant Dicke model \cite{Dicke54}.
Within our formalism, we can also address the criticality of the fully-connected Ising model imposing $\gamma=0$. 
This model presents, in the thermodynamical limit, a quantum phase transition due the spontaneous breaking of the $\mathbb{Z}_2$ symmetry \cite{Ising_infinite}.
 
The phase diagram of this model at zero temperature was derived in \cite{LMG_gaudin}, thanks to a two-boson Schwinger boson realization of the $SU(1,1)$ Richardson-Gaudin integrable models.  
However, classifying phase transitions in systems having finite number of elements 
is a challenging and an open problem. In particular, in the quantum domain, 
the issue relating to the scaling of the Hilbert space size, such us
the exponentially growing size of the Hilbert space for the considered systems, impacts strongly the performance of classical techniques.

We will study the precursors of the quantum phase transition for the finite size LMG model via quantum computational techniques. 
With abuse of notation, we will call the values of the magnetic field $B$ and of the anisotropy $\gamma$, for which 
the ground state and the first excited state of the system are degenerate, critical values. 

The adopted strategy can be easily extended to other critical systems, but the choice of the LMG model to test our approach is driven by two reasons.
On one side, the model is of interest for several communities and it has been used to test many-body approximations \cite{LMG_MB1,LMG_MB2}. 
The expression in Eq.\eqref{eq:LMGhamiltonian}, in terms of Pauli operators, makes the implementation 
on a superconducting quantum processor quite straightforward and requires less physical resources compared with its fermionic formulation.  
On the other side, the model has some peculiarities, namely anisotropy and long-range interaction, that makes it a non trivial model where to assess quantum criticality at finite-size. 

\section{Results}
This section presents the numerical results obtained in this work.
We remark  that a truly QPT is related to a singular behaviour of
the energy spectrum and the consequent non-analyticity of several observable quantities such as the magnetization. 
For systems of finite size a critical behaviour occurs when the ground-state energy has a level-crossing, 
viz., there is an interchange of the ground-state level and the first excited-state at a critical point.
This reflects in a null energy gap at the critical point and in a non-analytic behaviour of the mean magnetization along the model easy-axis.
Sec. \ref{sec:statevector} contains the ground and first excited state energy and magnetization for $N=4,5,6$ spins and different values of $\gamma$ and $B$ obtained on \cor{state vector} simulations and analyse their behaviour in the anisotropic and isotropic case. Sec. \ref{sec:hardware} shows the ground state energy and magnetization for $N=5$ spins, $\gamma=0.49$ and different values of $B$ computed on superconducting transmon qubits, and comments on the scalability of the VQE in the near term. 
We refer to \ref{sec:appendix} for a detailed explanation about different simulation backends and variational circuit optimizers.
\label{sec:results}

\subsection{Noiseless VQE Simulations}
\label{sec:statevector}
For a small number of spins $N$, the classical approach of the diagonalization of $H$ is straightforward. 
Defining the spectrum of the Hamiltonian $H \ket{\psi_n} = E_n \ket{\psi_n}$ with $n = \{0,..., 2^{N-1}\}$,
multiple crossing points between the energies of the ground and the first excited state ($E_{GS}$  and  $E_{1st}$, respectively) 
are found for the following critical values
\begin{equation}
    \label{eq:empiric}
    B_C^{k} = \frac{N - k}{N}\sqrt{\gamma},
\end{equation}
for a fixed $\gamma$, $k \leq N$ and odd \cite{LMG_MB2}.
Hence, the ground state energy can be recasted in $N/2 + 1$ \emph{phases}, if $N$ is even, or in $(N-1)/2 + 1$ otherwise. Introducing the ground-state magnetization along the model easy-axis $ \av{S_z} = \bra{\psi_0(B,\gamma)} S_z \ket{\psi_0(B,\gamma)}$, we can observe that for any nonanalytic point holds  $\lim_{B \to B_C^{k-}} \langle S_z \rangle \neq \lim_{B \to B_C^{k+}} \langle S_z \rangle \, \forall \, B_C^{(k)}$. 
In the following, we will denote $\ket{\psi_0} \equiv \ket{\psi_{GS}}$ and $\ket{\psi_1} \equiv \ket{\psi_{1st}}.$

We use \cor{state vector} simulations, as a theoretical tool to explore more complex ansatz and increase the performance as much as possible. We choose the depth of the ansatz as a function of the system size, namely $D=N $. This differs from the results obtained on real hardware, as explained in Sec. \ref{sec:ansatz}. The training is performed with the SLSQP optimizer \cite{kraft1988software} with 2000 maximum iterations. Adiabatic computing is applied to speed up the calculations and improve the accuracy, following the recommendation of \textcite{adiabatic_VQE}. 
The excited states are found using a VQD-like algorithm. Using the \cor{state vector} given by the solution of the Hamiltonian it is possible to redefine a new effective Hamiltonian
\begin{equation}
H'= H + \beta_{0}\ket{\psi_{GS}}\bra{\psi_{GS}}.
\end{equation}
When the superposition between $\ket{\psi_{1st}}$ found by using VQE on $H'$ and $\ket{\psi_{GS}}$ is small the loss function associated to this Hamiltonian reduces to Eq. (\ref{eq:VQD}) with $k=1$. We found this to be true every time the $\beta_0$ is set greater than the energy gap between the ground and first excited state, as specified in Sec.\ref{subsec:VQE}.
Knowing that transitions happen for $B=B_C^k$, see Eq. (\ref{eq:empiric}), we choose five points between the transition points and the chosen bounds. Starting from the upper bound, where the energy gap is wider, the ground state energy and the first excited energy are evaluated using random initial parameters. For each point in the interval, in decreasing order, the optimal point found in the previous step are chosen as initial parameters. Moreover, for the next interval, the optimal parameters for the ground state are used as initial point as well for the first excited state.  This technique significantly speeds up the simulation, improves the quality of the results, and allows us to compute the energies for systems up to $N=10$ spins, using \cor{state vector} simulations.

 \begin{figure}
\includegraphics[width=0.8\linewidth]{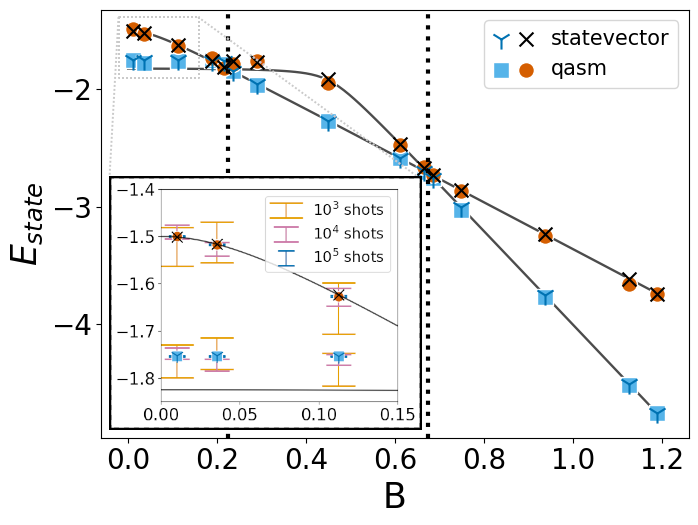}
\caption{\cor{We report the energies of the ground state $E_{GS}$ and first excited one $E_{1st}$ of the Hamiltonian in Eq. (\ref{eq:LMGhamiltonian}) for $N=4$ and for value of the anisotropy parameter $\gamma=0.81$ as a function of the magnetic field $B$. Solid lines represent the values obtained via exact classical diagonalization, while discrete points are the results obtained via VQE. The results for $E_{GS}$ are marked in light blue with tri-down markers (state vector simulation) and filled squares (QASM noiseless simulation), while results for $E_{1st}$ are in orange with crosses marking the state vector simulation and filled circles the QASM noiseless outputs.  
Vertical dotted lines show crossing points for values given in Eq. \eqref{eq:empiric}. The inset makes it possible to better appreciate the accuracy of energy estimates as a function of the number of shoots, where the error bar correspond to one standard deviation.}}
\label{fig:E_4at}   
\end{figure}
        
\begin{figure}
\includegraphics[width=0.8\linewidth]{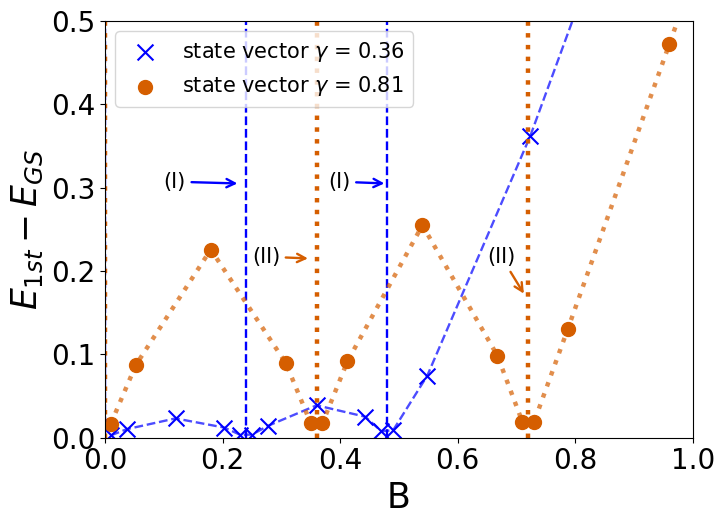}
\caption{\cor{We plot the difference between the energy of the first excited state $E_{1st}$ and the ground state energy $E_{GS}$ as a function of the tuning field $B$. The results are shown for a system with $N=5$ lattice sites, focusing on different values of the anisotropy parameter $\gamma = 0.36, \, 0.81$. The results obtained via VQE are marked by blue crosses ($\gamma = 0.36$) and orange dots ($\gamma = 0.81$). The blue dashed line ($\gamma = 0.36$) and the orange dotted line ($\gamma = 0.81$) represent the exact diagonalization values of $E_{1st}- E_{GS}$. Vertical lines correspond to values of the critical magnetic field such that Eq. (\ref{eq:empiric}) holds, where the label (I) refers to $\gamma = 0.36$, and label (II) to $\gamma = 0.81$).}}
\label{fig:gap_gammas}  
\end{figure}

\begin{figure}
\includegraphics[width=0.8\linewidth]{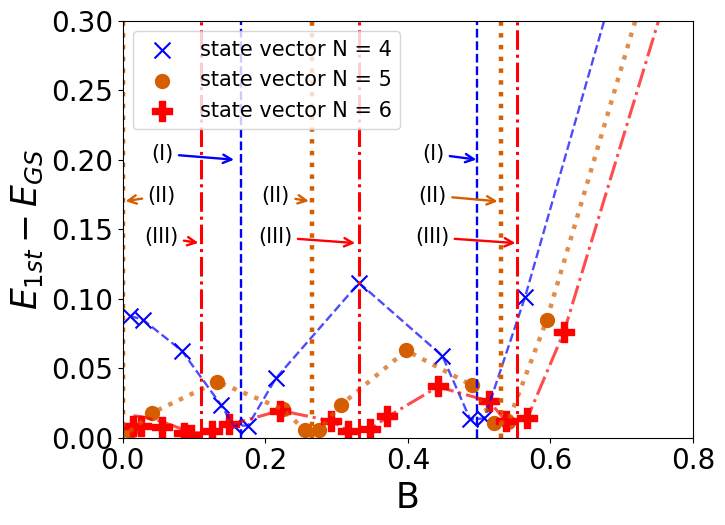}
\caption{\cor{We report the energy difference $E_{1st}- E_{GS}$ as a function of the magnetic field $B$. The results are computed  at $\gamma = 0.44$ for different lattice sizes $N=4$ (blue), $N=5$ (orange), $N=6$ (red).
The results obtained via VQE are represented by a blue cross ($N=4$), an orange dot ($N=5$) and a red plus ($N=6$), while the blue dashed line ($N=4$), the orange dotted line ($N=5$), and a red dotdashed line ($N=6$) represent the values obtained by exact diagonalization of the Hamiltonian $H$. The vertical lines correspond to the values of the magnetic field for which  by we observe a crossing points between the two lowest state  energies (see Eq. (\ref{eq:empiric}), they are labelled with (I) for $N = 4$, (II) for $N = 5$ and (III) for $N=6$ sites.}}
\label{fig:gap_atoms}   
\end{figure}

Figure \ref{fig:E_4at} shows the ground and first excited state energy for a LMG Hamiltonian with $N=4$ spins for the specific value of $\gamma = 0.81$, as a function of the magnetic field $B$. The VQE is compared to the  exact diagonalization (black solid lines), while the VQE points are obtained using \cor{state vector} simulation (crosses) as well as shoots-based probabilistic output without hardware noise contribution (\cor{filled squares and circles}).
Another figure of merit to assess a quantum phase transition for finite-size systems is the energy difference between the first excited state and the ground state, namely $E_{1st} - E_{GS}$ (to which we will refer to as the gap, for shorthand of notation).
The gap as a function of the magnetic field  is shown in Figure \ref{fig:gap_gammas} for $N=5$ spins and two values of $\gamma$, $\gamma = 0.36,0.81$. Even if they are far from the extreme values $0,1$ they already underline a difference in the behaviour, at least for $B<0.6$. Similar considerations hold as a function of the system size, shown in  Figure \ref{fig:gap_atoms} for $N=4,5,6$ with \textcolor{black}{$\gamma=0.44$}. In both cases, the energy gap rapidly explodes after $B \geq0.6$. 

\begin{figure*}
            \includegraphics[width=0.65\columnwidth]{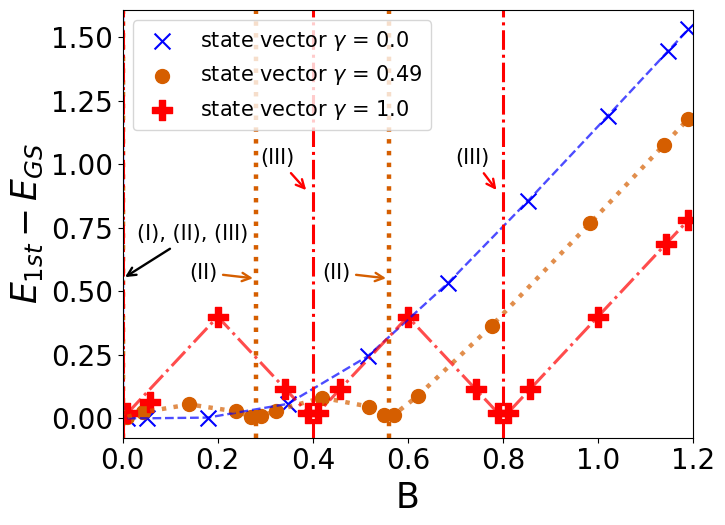}
            \includegraphics[width=0.65\columnwidth]{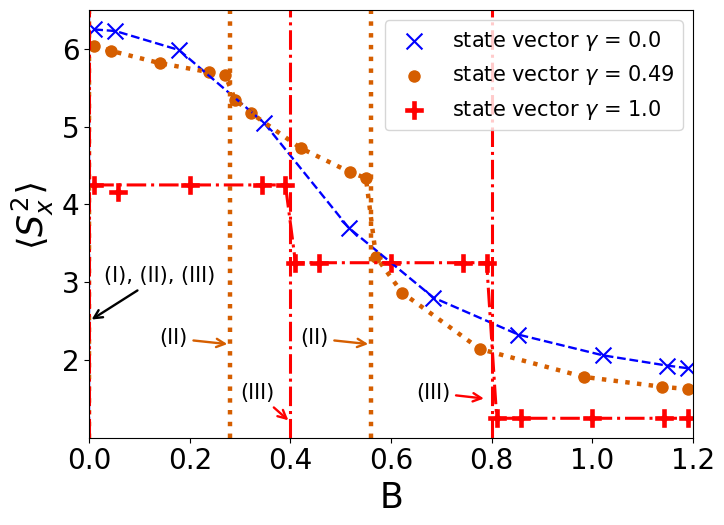}
            \includegraphics[width=0.65\columnwidth]{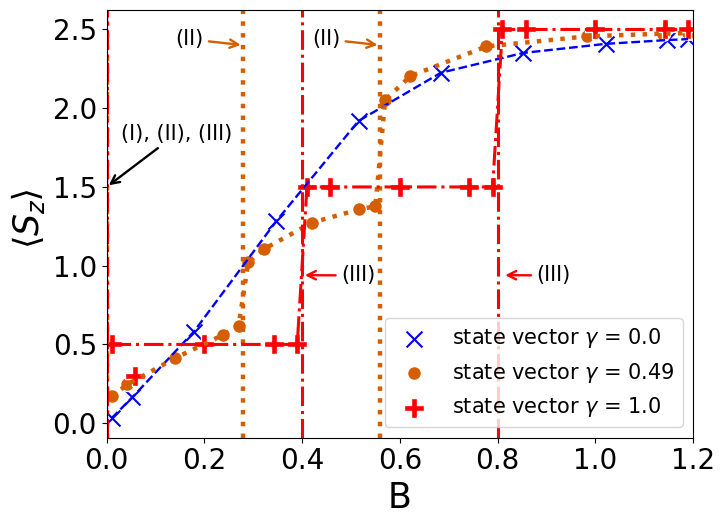}
\caption{\cor{We report the three relevant quantities to detect the criticality in a fully connected spin system. In (a) we show the energy gap  $E_{1st}- E_{GS}$. In (b) we plot the correlation function $\left<S_x^2\right>$, and in (c) the mean magnetization along the direction of the magnetic field $\langle S_{z} \rangle$. The results are shown for $N=5$ spins for particular values of the anisotropy identifying different classes of models $\gamma = 0$ (fully connected Ising model), $\gamma = 0.49$ (LMG model), and $\gamma = 1$ (Lipkin-Dicke model). In all the three plots, the results obtained via the quantum variational algorithm are represented by blue crosses ($\gamma = 0$), orange dots ($\gamma = 0.49$), and red plusses ($\gamma = 1$) and the benchmarking is given by the corresponding exact diagonalization values, shown as a blue dashed line ($\gamma = 0$), an orange dotted line ($\gamma = 0.49$), and a red dotdashed line ($\gamma = 1$). The results are in agreement with the critical values in Eq. (\ref{eq:empiric}) and such values are reported as vertical lines (label (I) points the only trivial value for $\gamma = 0$, (II) points critical values for $\gamma = 0.49$, and (III) for $\gamma = 1$).}}
\label{fig:gapS2xSz_extreme}
\end{figure*}

Finally, we consider the extreme cases of the fully-connected Ising model and the Lipkin-Dicke model, for $\gamma = \{0,1 \}$ respectively. We report our results in Figure \ref{fig:gapS2xSz_extreme}, together with an intermediate $\gamma$ value of $0.49$, where the system size is fixed to $N=5$. The energy gap between the first excited state and the ground state is shown in (a), while the correlation function along the $x-$axis $\left<S_x^2\right>$ in (b) and the magnetization along the longitudinal direction of the magnetic field $\langle S_{z} \rangle$ in (c). \cor{We have decided to plot the correlation function $\left<S_x^2\right>$ because, due to the spin-flip simmetry of the Hamiltonian in Eq. (\ref{eq:LMGhamiltonian}), the mean value of the magnetization along the plane perpendicular the magnetic field is zero (see \cite{LMG_MB1}.)}  We observe the oscillatory trend of the energy gap for $\gamma  = 1$ as opposed to the monotonic trend of the isotropic case ($\gamma=0$). A completely different behaviour can be appreciated also for the two magnetization observable, where the anisotropic model is characterised by a stepwise trend as opposed to the continuous one for $\gamma  = 0$, signaling how the three models in the thermodynamical limit belong to distinct universality classes. 

\subsection{Runs on the Real Devices.}
\label{sec:hardware}
\subsubsection{Experimental Device}
\label{hardware_description}
\begin{figure}
    \centering
    \includegraphics[scale=0.3]{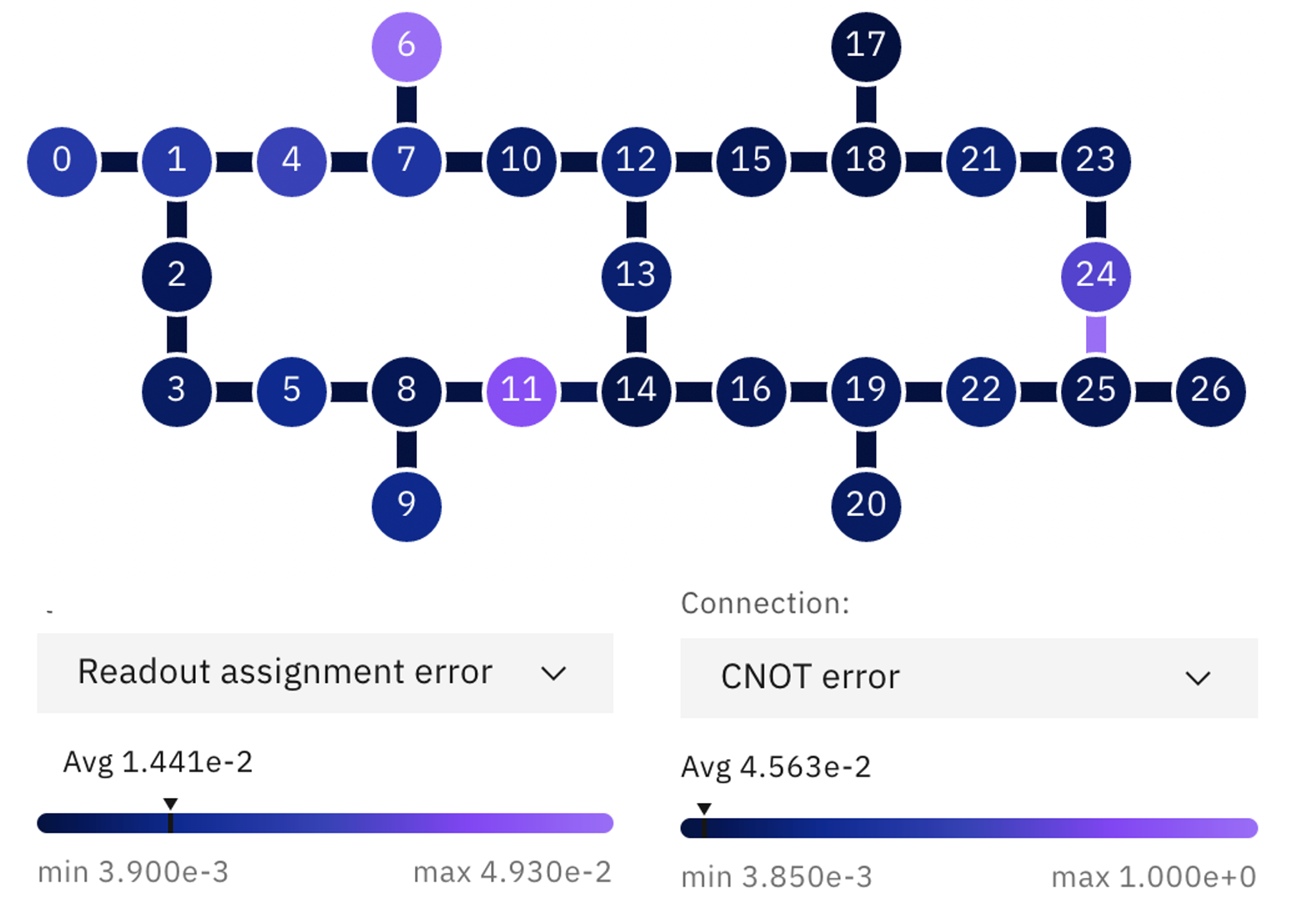}
    \caption{Topology of the superconducting quantum device ibmq\_kolkata with color map representation. Color associated to each qubit represents the readout error at the time of calibration while the color of the connection between two qubits represents the CNOT error rate. Image taken from the IBM Q Lab \url{https://quantum-computing.ibm.com/services/resources?system=ibmq_kolkata}}
    \label{fig:topology}
\end{figure}
The quantum device used in this work consists of 27 fixed\hyp frequency transmons qubits, with fundamental transition frequencies of approximately 5 GHz and anharmonicities of $-340$ MHz, with the same topology as displayed in Figure \ref{fig:topology}. Microwave pulses are used for single\hyp qubit gates and cross\hyp resonance interaction \cite{cross-resonance} for two\hyp qubit gates. The experiments took place over one month, but each different computation took place over span of five hours, without intermediate calibration, with the use of Qiskit Runtime. The median qubit lifetime $T_1$ of the qubits is 121 and 129 $\mu s$, the median coherence time  $T_2$ is 90 and 135 $\mu s$ and the median readout and CNOT error is 0.014 and 0.045 respectively. The SABRE \cite{sabre} algorithm is used for the transpilation to the quantum hardware.

\subsubsection{Small System Size}
\begin{figure*}
        \includegraphics[width=0.65 \columnwidth]{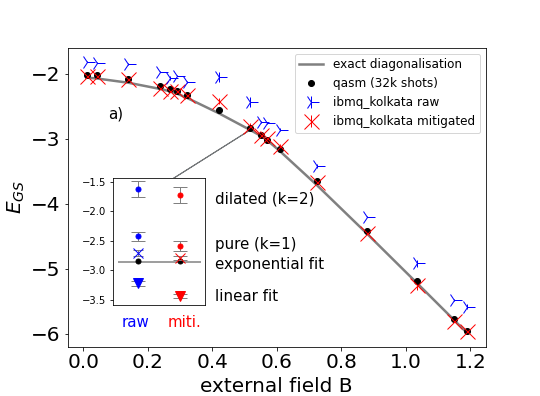}
        \includegraphics[width= 0.65\columnwidth]{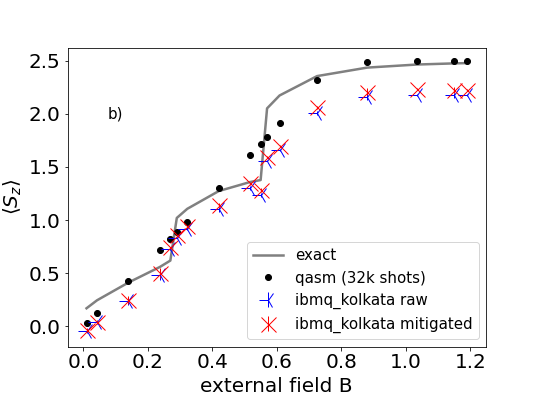}
        \includegraphics[width=0.65 \columnwidth]{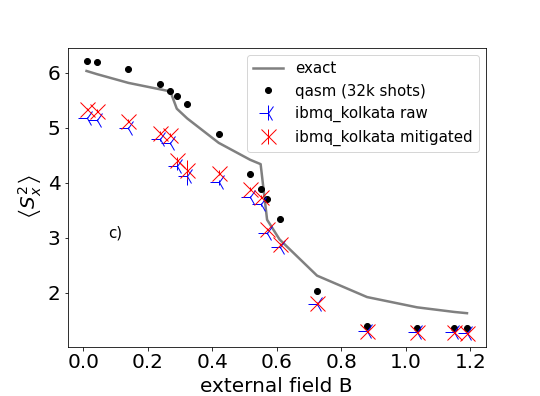}
\caption{\cor{We report the three relevant quantities to detect the criticality in a fully connected spin system, using the VQE algorithm run on the superconducting device ibmq\_kolkata. The energy gap  $E_{1st}- E_{GS}$ is in (a). The mean value of two magnetic observables: in (b) the correlation function along an axis transverse to the magnetic field $\left<S_x^2\right>$, and in (c) the mean magnetization along the direction of the magnetic field $\langle S_{z} \rangle$. The results are done for a lattice size of $N=5$ spins at $\gamma = 0.49$.
Points are obtained on the superconducting device ibmq\_kolkata with (red crosses) and without (blue tri-left markers) error mitigation. 
The experimental values are compared to noiseless simulation (black dots) and the final benchmark is given by the exact diagonalization values (solid line). The inset in (a) shows the extrapolation to the zero noise regime, both with an exponential and linear fit.}}
\label{fig:hardware_5}
\end{figure*}

We begin by computing the ground state energy of a system with $N=5$ spins and $\gamma=0.49$ for different values of the magnetic field $B$. We use the hardware efficient ansatz with $D=1$ repetition, as shown in Figure \ref{fig:ansatz}. As a warm initialization, the ansatz is first trained on the noiseless simulator, and the optimal parameters are used as an educated guess for the initial parameters. The training is composed of maximum 100 steps, or until convergence, with the SPSA \cite{SPSA} optimizer using a learning rate of 0.005 for the first 30 steps and 0.001 afterwards, using 8092 shots. The graphs are obtained with 32000 shots and statistics are collected from 5 distinct runs. Measurement error mitigation and zero noise extrapolation are performed to enhance the results, which are shown in Figure \ref{fig:hardware_5}. Solid lines correspond to the exact diagonalization, the black dots to the noiseless qasm simulation with 32000 shots, the blue crosses to the raw results from \cor{the quantum device} and the red ones to the mitigated energies. The error bars correspond to the 99.5 \% confidence interval. The inset shows the effect of \cor{different} error mitigation \cor{tuning} on a specific point. The $k=1$ point correspond to the original circuit while $k=2$ to the dilated case, where every CNOT is replaced with three CNOTs. The cross shows a scaled exponential fit while the triangle a linear one. As motivated in Sec. \ref{sec:mitigation}, the linear fit overshoots the true ground state energy, while this is not the case for the scaled exponential fit.

We observe that the ground state energy is reproduced with less than 1\% error ratio everywhere, suggesting that the quality of current devices is good enough for such tasks. However, the computed magnetization observables are not equally accurate. The explanation is two folds: first, we observe that the noiseless simulations are also less precise than the energy calculations, more particularly for $\left<S_x^2\right>$ at large magnetic field. This is essentially caused by the ansatz which is too shallow to represent the true ground state, but instead is only a good approximation with similar energy. But more importantly, there is a discrepancy between the noiseless and real hardware results, which is due to overfitting to the hardware noise. 
For the ground state calculation the noise is adapted to get to the GS, according to the real condition, which includes the presence of noise, and this is why we refer to this as an overfitting behaviour.
By doing so, we get closer to the true energy, but drift from the true ground state. 
The approximation in computing the correct ground state is amplified when $\left<S_x^2\right>$ and $\left<S_z\right>$ are computed shifting uniformly the curve.

\subsubsection{Discussion on Large System Size}
Finally, we tried to extend the reach of VQE to sizes where simulations are unavailable due to the exponential scaling of the Hilbert space. Even if density\hyp matrix renormalization group (DMRG) \cite{DMRG} techniques are able to compute the ground state energy for large number of spins ($\sim 10^2$), we choose $N=20$ since \cor{it is out of reach, in terms of simulation time, for our availability.}
 This problem is more interesting than the previous case since we are unable to start from a set of previously trained parameters. \cor{In addition}, these calculations are also more challenging for current devices for the following reasons. Gradient\hyp free optimizers, such as SPSA, require small amount of circuit executions to estimate the gradient. \cor{Yet}, since they rely on finite difference techniques, the gradient is strongly affected by the noise and can lead to erratic path in the optimization landscape. On the other hand, analytical gradients provided by the parameter\hyp shift rule \cite{grad} are more reliable, but also more expensive to compute since they require $2\cdot d$ circuit executions, where $d$ is the number of parameters ($d=40$ in this case). \cor{Accordingly, we estimate} more than one hour runtime per optimization step, accounting for \cor{error mitigation techniques} e.g. ZNE and MEM, which is more than what we can reasonably obtain from on a shared device and without running into further recalibration problems \cite{ZEN_Takagi, ZEN_Endo}.
This is one of the main reason why innovative integrated architecture of quantum and classic computer like the one proposed by IBM with \href{https://research.ibm.com/blog/qiskit-runtime-for-useful-quantum-computing}{Qiskit Runtime} would strongly reduce the computation time.
\section{Higher excited states}\label{sec:excited}
\begin{figure}
   \includegraphics[width=0.8\linewidth]{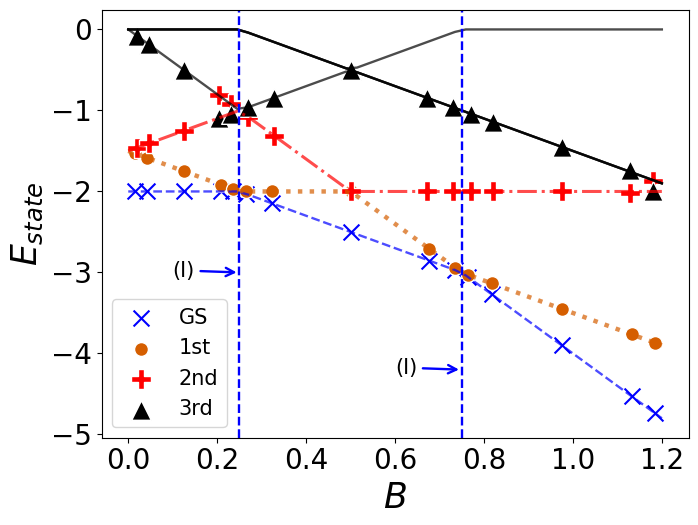}
\caption{\textcolor{black}{Representation of various excited states for $N=4$ and $\gamma=1$. VQE results are represented by a blue cross (ground-state), an orange dot (1st excited state), a red plus (2nd excited state), and a black triangle (3rd excited state).
The blue dashed line (ground state), the orange dotted line (1st excited state), the red dotdashed line (2nd excited state), and the black solid lines (3rd, 4th, 5th and 6th excited states, some are degenerate) represent the classical diagonalization values. Vertical lines indicate crossing points.}}
\label{fig:excited_g1}
\end{figure}

\begin{figure}[t!]
   \includegraphics[width=0.8\linewidth]{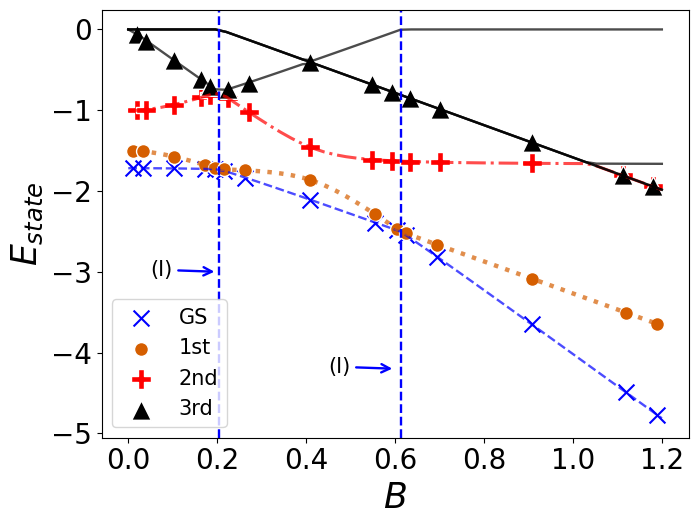}
\caption{\cor{We plot the energies $E_{2nd}$ and $E_{3rd}$ of the two excited states, beyond the first one, for the Hamiltonian in Eq. (\ref{eq:LMGhamiltonian}) with $N=4$ at the value anisotropy $\gamma=0.67$. In order to make the comparison clear, we report also the values of the energies of the ground state $E_{GS}$ and of the first one $E_{1st}$. The markers of the VQE results are the following: blue cross for the ground-state energy, an orange dot for the first excited state, a red plus for second excited state, and a black triangle  for third excited state.
Exact diagonalization results are shown as a blue dashed line (ground state), an orange dotted line (first excited state), a red dotdashed line (second excited state), while the black solid lines are for the energies of degenerated states (ranging from the third excited and up to the sixth) Vertical lines indicate crossing points (see Eq. (\ref{eq:empiric}).}}
\label{fig:excited_g}
\end{figure}
The VQE can be used to compute the energies of the second and third excited states as well. Figures \ref{fig:excited_g1} and \ref{fig:excited_g} show the \textcolor{black}{seven} lowest energy eigenvalues for a system of $N=4$ spins, as a function of the magnetic field $B$ at two different interaction configurations $\gamma= \{1, 0.67\}$, respectively. The simulations are performed using \cor{state vector} and superimposed to exact diagonalization. For $\gamma = 0.67$, VQE seems at first to fail in computing the 3rd excited state, but actually \textcolor{black}{finds} degenerate states. To better understand the degeneracy, let us consider $\gamma=1$ and use Eq. (\ref{eq:LMGDickeHamiltonian}) to obtain
\begin{equation}
    H\ket{j,m} = \bigl[ -\frac{1}{2}(j(j+1) - m^2 - 2) - 2Bm\bigr]\ket{j,m}.
\end{equation}
For $N=4$ spins, $j = 0, 1, 2$, leading to $9$ distinct degenerate values for the energies of the 16 eigenstates.  
The first degenerate eigenvalue  for $B < \frac 1 4$, is the one with the $j=1, \, m=1$ quantum numbers. However, it becomes the fourth excited for $\frac 1 4 < B < \frac 1 2$, the third for $\frac 1 2 < B < \frac 5 4$, and finally the second  for $B$ greater than $\frac 5 4$.
Numerical investigations suggest an analogous behaviour for $\gamma \neq 1$ and $0<\gamma\lesssim 2$. In the region $B \lesssim \frac{5}{4}\sqrt{\gamma}$, the first eigenvalue to be degenerate is the third excited while in $B \gtrsim \frac{5}{4}\sqrt{\gamma}$, it is the second excited. Hence, VQE actually shows  in Figure \ref{fig:excited_g} that for $B \approx 1.2$, the third eigenvalue is degenerate (3 times, in particular).
The numerical investigations for $\gamma=1$ show that the degenerate levels are
\begin{itemize}
    \item $3$-fold: $j=1$, $m\in\{0,\pm1\}$,
    \item $2$-fold: $j=0$, $m=0$,
\end{itemize}
in agreement with \cite{Dicke54}.
A similar argument can be addressed also to justify the behaviour of the energies of the excited states for the model with $\gamma \neq 1 $ as those observed in Figure \ref{fig:excited_g}. However, the impossibility to diagonalize the two terms of the Hamiltonian in a common basis would make the argument only less intuitive and more cumbersome.

\section{Discussion and Outlooks}
\label{sec:discussion}

The advent of reliable quantum hardware, although not yet fault-tolerant, 
has paved the way to novel techniques to tackle problems from different research areas. 
A natural avenue of research is the one that incorporates the quantum computing techniques 
to understand the physics of complex systems as many-body systems. 
To this end, we have proposed a way to exploit the variational quantum eigensolver, 
and the algorithms stemmed from it, for studying the finite-size criticality of paradigmatic spin models. Upon introducing a Hamiltonian with an anisotropic interaction $\gamma$ in the $x-y$ 
plane we have studied the level crossings 
between the ground-state and the first-excited state in the proper LMG ($0< \gamma < 1)$ and in the two limit cases: the fully connected Ising model ($\gamma =0$) and the Lipkin-Dicke model $\gamma=1.$  
We used as figure of merit some relevant magnetic observables, i.e., 
the magnetization along the field direction and the spin-spin correlation along the $x-$axis. 

Due to the geometry of the system, no length scale of the correlation can be defined, this makes the fully-connected spin models interesting systems to look for unconventional results at finite-size \cite{LMG_MB1, LMG_MB2} or to give a quantitative evaluation of the quality of a new computational or experimental technique \cite{LMG4, LMG_impl1, LMG_QT1}. 

Looking at this LMG model the number of measurements scale maximally at three, for the three independent terms in the hamiltonian that do not commute. According to the results presented so far, it turns out that a limiting factor in getting better performance is the noise while barren plateau represents potentially a second order factor.

Recently, several papers \cite{LMG_QC1,LMG_QC2,LMG_QC3,LMG_QC4} addressed the Lipkin model on a quantum computer to question whether techniques and methods proper of quantum machine learning can be employed in nuclear physics. 
In general, the authors rely on system size of relatively small dimensions $N\leq 4$ 
performing a preliminary noiseless analysis for the isotropic model that can be analytically solved exactly. 

Our analysis is complementary to those previously carried out and go into the direction of employing quantum algorithms to have a direct insight on problems of relevance in statistical physics. 
In fact, we have shown that the VQE is a powerful tool to assess the quantum phase transition of critical systems of finite-size. We have also addressed how to mitigate the errors present when employing NISQ devices and how it is feasible on a real hardware based on superconducting transmon qubits. 

As final remark and open question we surmise that our method could be employed in future, when better performing hardware, with more qubits will be available, as a benchmark for the renormalization group approaches used to study 
the finite-size scaling behaviour of quantities of interest in statistical and condensed matter theory.

\section*{Acknowledgments}
AM is supported by Foundation for Polish Science (FNP), IRAP project ICTQT, contract no. 2018/MAB/5, co-financed by EU  Smart Growth Operational Programme, and (Polish) National Science Center (NCN), MINIATURA  DEC-2020/04/X/ST2/01794. MG and OK are supported by CERN Quantum Technology Initiative.
Access to the IBM Quantum Services was obtained through the IBM Quantum Hub at CERN. The views expressed are those of the authors and do not reflect the official policy or position of IBM or the IBM~Q team. The authors thank Professor Angelo Bassi and the University of Trieste for the hospitality. 
\appendix*

\section{Performance of the optimizer}
\label{sec:appendix}
\begin{figure*}
        \includegraphics[width=1.75 \columnwidth]{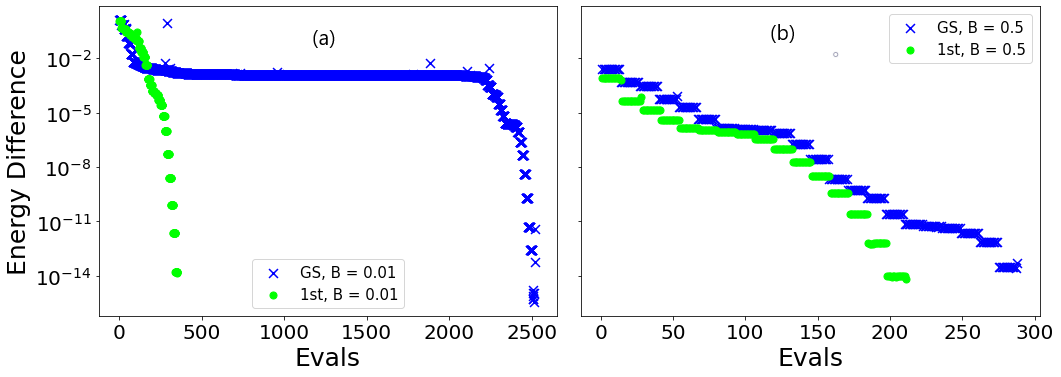}
     \caption{\textcolor{black}{We plot the  SLSQP loss curve for the two lowest energy states for different values of the magnetic field. Namely, in (a) we set $B = 0.01,$ and in (b) $B = 0.5$ (b), for $N=4$ spins at $\gamma = 0.49$. The blue crosses refers to the ground state (GS) and the green filled circles to the first excited state (1st).}}
\label{fig:loss_curve}
\end{figure*}

In this section we give additional information and further explanation about the techniques adopted in this work. The gate\hyp based quantum circuits used in this work are built using the open\hyp source framework \href{https://qiskit.org/documentation/nature/}{\texttt{qiskit-nature}} \cite{Qiskit}.

The \textit{state vector simulation} results are referred to classical simulation of a quantum circuit, probing the potential of this approach under ideal conditions, such as using exponentially many shots, without noise. 
For the optimization in a \cor{state vector} setting we use SLSQP \cite{slsqp}, Sequential Least SQuares Programming. It minimizes a function of several variables with any combination of bounds, equality and inequality constraints. We refer to \href{https://docs.scipy.org/doc/scipy/reference/generated/scipy.optimize.minimize.html}{scipy} for further details.

\textit{QASM} results are still referred to a full classical simulation of a quantum circuit plus measurement. This implies the simulation of probability outcomes of quantum observable given as a sequence of bit strings with the relative counts or number of repetitions.  For the optimization in a \cor{state vector} setting we use COBYLA \cite{adiabatic_VQE}. This is a gradient-free optimizers, likewise the one used for optimization on the quantum computer: the simultaneous perturbation stochastic approximation (SPSA)~\cite{SPSA} optimizer. Differently from gradient based optimization, where the next best parameters in the optimization is obtained from the gradient of a given function with a high possibility to be stuck in a local minima$/$maxima when traversing parameter(s), 
SPSA efficiently approximates the gradient with few circuit evaluations by shifting the parameters in two random directions. The learning rate is changed at every bunch of iterations to ensure a fast convergence at the beginning and avoid oscillations at the end. Looking at realistic experiments, the stochastic nature of SPSA makes it resilient to the statistical noise coming from the finite number of measurements, making it appealing for quantum devices.
A quantum natural variant of the SPSA optimizer using the geometry of the Hilbert space has been recently proposed by ~\textcite{QNSPSA}. In this variant the Hessian is approximated with six circuit evaluations and significantly improves the optimization efficiency of quantum circuits. In the present work, this optimizer is not considered.

To better understand the performance and the behaviour of the optimizer we provide the loss curve during the learning process for the SLSQP (state vector simulation) optimizer in Fig.\ref{fig:loss_curve}. The energy difference in the plot is given by: 
\begin{equation}
    \log{\frac{E_{count} - E_{final}}{E_{final}}}. 
\end{equation}
The huge difference in the number of evaluations needed between the first $B$ value and $B = 0.5$ shows one of the benefits of taking the last $B$ value optimal parameters as the initial ones for the next $B$.


%


\begin{thebibliography}{92}%
\makeatletter
\providecommand \@ifxundefined [1]{%
 \@ifx{#1\undefined}
}%
\providecommand \@ifnum [1]{%
 \ifnum #1\expandafter \@firstoftwo
 \else \expandafter \@secondoftwo
 \fi
}%
\providecommand \@ifx [1]{%
 \ifx #1\expandafter \@firstoftwo
 \else \expandafter \@secondoftwo
 \fi
}%
\providecommand \natexlab [1]{#1}%
\providecommand \enquote  [1]{``#1''}%
\providecommand \bibnamefont  [1]{#1}%
\providecommand \bibfnamefont [1]{#1}%
\providecommand \citenamefont [1]{#1}%
\providecommand \href@noop [0]{\@secondoftwo}%
\providecommand \href [0]{\begingroup \@sanitize@url \@href}%
\providecommand \@href[1]{\@@startlink{#1}\@@href}%
\providecommand \@@href[1]{\endgroup#1\@@endlink}%
\providecommand \@sanitize@url [0]{\catcode `\\12\catcode `\$12\catcode
  `\&12\catcode `\#12\catcode `\^12\catcode `\_12\catcode `\%12\relax}%
\providecommand \@@startlink[1]{}%
\providecommand \@@endlink[0]{}%
\providecommand \url  [0]{\begingroup\@sanitize@url \@url }%
\providecommand \@url [1]{\endgroup\@href {#1}{\urlprefix }}%
\providecommand \urlprefix  [0]{URL }%
\providecommand \Eprint [0]{\href }%
\providecommand \doibase [0]{http://dx.doi.org/}%
\providecommand \selectlanguage [0]{\@gobble}%
\providecommand \bibinfo  [0]{\@secondoftwo}%
\providecommand \bibfield  [0]{\@secondoftwo}%
\providecommand \translation [1]{[#1]}%
\providecommand \BibitemOpen [0]{}%
\providecommand \bibitemStop [0]{}%
\providecommand \bibitemNoStop [0]{.\EOS\space}%
\providecommand \EOS [0]{\spacefactor3000\relax}%
\providecommand \BibitemShut  [1]{\csname bibitem#1\endcsname}%
\let\auto@bib@innerbib\@empty
\bibitem [{\citenamefont {Zinn-Justin}(2007)}]{zinn2007phase}%
  \BibitemOpen
  \bibfield  {author} {\bibinfo {author} {\bibfnamefont {Jean}\ \bibnamefont
  {Zinn-Justin}},\ }\href@noop {} {\emph {\bibinfo {title} {Phase transitions
  and renormalization group}}}\ (\bibinfo  {publisher} {Oxford University Press
  on Demand},\ \bibinfo {year} {2007})\BibitemShut {NoStop}%
\bibitem [{\citenamefont {Pelissetto}\ and\ \citenamefont
  {Vicari}(2002)}]{pelissetto2002critical}%
  \BibitemOpen
  \bibfield  {author} {\bibinfo {author} {\bibfnamefont {Andrea}\ \bibnamefont
  {Pelissetto}}\ and\ \bibinfo {author} {\bibfnamefont {Ettore}\ \bibnamefont
  {Vicari}},\ }\bibfield  {title} {\enquote {\bibinfo {title} {Critical
  phenomena and renormalization-group theory},}\ }\href@noop {} {\bibfield
  {journal} {\bibinfo  {journal} {Physics Reports}\ }\textbf {\bibinfo {volume}
  {368}},\ \bibinfo {pages} {549--727} (\bibinfo {year} {2002})}\BibitemShut
  {NoStop}%
\bibitem [{\citenamefont {Englert}\ and\ \citenamefont
  {Brout}(1964)}]{higgsmechanism2}%
  \BibitemOpen
  \bibfield  {author} {\bibinfo {author} {\bibfnamefont {F.}~\bibnamefont
  {Englert}}\ and\ \bibinfo {author} {\bibfnamefont {R.}~\bibnamefont
  {Brout}},\ }\bibfield  {title} {\enquote {\bibinfo {title} {Broken symmetry
  and the mass of gauge vector mesons},}\ }\href {\doibase
  10.1103/PhysRevLett.13.321} {\bibfield  {journal} {\bibinfo  {journal} {Phys.
  Rev. Lett.}\ }\textbf {\bibinfo {volume} {13}},\ \bibinfo {pages} {321--323}
  (\bibinfo {year} {1964})}\BibitemShut {NoStop}%
\bibitem [{\citenamefont {Higgs}(1964)}]{higgsmechanism}%
  \BibitemOpen
  \bibfield  {author} {\bibinfo {author} {\bibfnamefont {Peter~W.}\
  \bibnamefont {Higgs}},\ }\bibfield  {title} {\enquote {\bibinfo {title}
  {Broken symmetries and the masses of gauge bosons},}\ }\href {\doibase
  10.1103/PhysRevLett.13.508} {\bibfield  {journal} {\bibinfo  {journal} {Phys.
  Rev. Lett.}\ }\textbf {\bibinfo {volume} {13}},\ \bibinfo {pages} {508--509}
  (\bibinfo {year} {1964})}\BibitemShut {NoStop}%
\bibitem [{\citenamefont {Bardeen}\ \emph {et~al.}(1957)\citenamefont
  {Bardeen}, \citenamefont {Cooper},\ and\ \citenamefont {Schrieffer}}]{BCS}%
  \BibitemOpen
  \bibfield  {author} {\bibinfo {author} {\bibfnamefont {J.}~\bibnamefont
  {Bardeen}}, \bibinfo {author} {\bibfnamefont {L.~N.}\ \bibnamefont {Cooper}},
  \ and\ \bibinfo {author} {\bibfnamefont {J.~R.}\ \bibnamefont {Schrieffer}},\
  }\bibfield  {title} {\enquote {\bibinfo {title} {Theory of
  superconductivity},}\ }\href {\doibase 10.1103/PhysRev.108.1175} {\bibfield
  {journal} {\bibinfo  {journal} {Phys. Rev.}\ }\textbf {\bibinfo {volume}
  {108}},\ \bibinfo {pages} {1175--1204} (\bibinfo {year} {1957})}\BibitemShut
  {NoStop}%
\bibitem [{\citenamefont {Tilley}\ and\ \citenamefont
  {Tilley}(2019)}]{tilley2019superfluidity}%
  \BibitemOpen
  \bibfield  {author} {\bibinfo {author} {\bibfnamefont {David~R}\ \bibnamefont
  {Tilley}}\ and\ \bibinfo {author} {\bibfnamefont {John}\ \bibnamefont
  {Tilley}},\ }\href@noop {} {\emph {\bibinfo {title} {Superfluidity and
  superconductivity}}}\ (\bibinfo  {publisher} {Routledge},\ \bibinfo {year}
  {2019})\BibitemShut {NoStop}%
\bibitem [{\citenamefont {Amico}\ \emph {et~al.}(2008)\citenamefont {Amico},
  \citenamefont {Fazio}, \citenamefont {Osterloh},\ and\ \citenamefont
  {Vedral}}]{manybody_rev}%
  \BibitemOpen
  \bibfield  {author} {\bibinfo {author} {\bibfnamefont {Luigi}\ \bibnamefont
  {Amico}}, \bibinfo {author} {\bibfnamefont {Rosario}\ \bibnamefont {Fazio}},
  \bibinfo {author} {\bibfnamefont {Andreas}\ \bibnamefont {Osterloh}}, \ and\
  \bibinfo {author} {\bibfnamefont {Vlatko}\ \bibnamefont {Vedral}},\
  }\bibfield  {title} {\enquote {\bibinfo {title} {Entanglement in many-body
  systems},}\ }\href {\doibase 10.1103/RevModPhys.80.517} {\bibfield  {journal}
  {\bibinfo  {journal} {Rev. Mod. Phys.}\ }\textbf {\bibinfo {volume} {80}},\
  \bibinfo {pages} {517--576} (\bibinfo {year} {2008})}\BibitemShut {NoStop}%
\bibitem [{\citenamefont {Sch{\"a}fer}\ \emph {et~al.}(2020)\citenamefont
  {Sch{\"a}fer}, \citenamefont {Fukuhara}, \citenamefont {Sugawa},
  \citenamefont {Takasu},\ and\ \citenamefont {Takahashi}}]{manybody_rev2}%
  \BibitemOpen
  \bibfield  {author} {\bibinfo {author} {\bibfnamefont {Florian}\ \bibnamefont
  {Sch{\"a}fer}}, \bibinfo {author} {\bibfnamefont {Takeshi}\ \bibnamefont
  {Fukuhara}}, \bibinfo {author} {\bibfnamefont {Seiji}\ \bibnamefont
  {Sugawa}}, \bibinfo {author} {\bibfnamefont {Yosuke}\ \bibnamefont {Takasu}},
  \ and\ \bibinfo {author} {\bibfnamefont {Yoshiro}\ \bibnamefont
  {Takahashi}},\ }\bibfield  {title} {\enquote {\bibinfo {title} {Tools for
  quantum simulation with ultracold atoms in optical lattices},}\ }\href@noop
  {} {\bibfield  {journal} {\bibinfo  {journal} {Nature Reviews Physics}\
  }\textbf {\bibinfo {volume} {2}},\ \bibinfo {pages} {411--425} (\bibinfo
  {year} {2020})}\BibitemShut {NoStop}%
\bibitem [{\citenamefont {Sachdev}(2011)}]{sachdev_2011}%
  \BibitemOpen
  \bibfield  {author} {\bibinfo {author} {\bibfnamefont {Subir}\ \bibnamefont
  {Sachdev}},\ }\href {\doibase 10.1017/CBO9780511973765} {\emph {\bibinfo
  {title} {Quantum Phase Transitions}}},\ \bibinfo {edition} {2nd}\ ed.\
  (\bibinfo  {publisher} {Cambridge University Press},\ \bibinfo {year}
  {2011})\BibitemShut {NoStop}%
\bibitem [{\citenamefont {Sondhi}\ \emph {et~al.}(1997)\citenamefont {Sondhi},
  \citenamefont {Girvin}, \citenamefont {Carini},\ and\ \citenamefont
  {Shahar}}]{RevQPT}%
  \BibitemOpen
  \bibfield  {author} {\bibinfo {author} {\bibfnamefont {S.~L.}\ \bibnamefont
  {Sondhi}}, \bibinfo {author} {\bibfnamefont {S.~M.}\ \bibnamefont {Girvin}},
  \bibinfo {author} {\bibfnamefont {J.~P.}\ \bibnamefont {Carini}}, \ and\
  \bibinfo {author} {\bibfnamefont {D.}~\bibnamefont {Shahar}},\ }\bibfield
  {title} {\enquote {\bibinfo {title} {Continuous quantum phase transitions},}\
  }\href {\doibase 10.1103/RevModPhys.69.315} {\bibfield  {journal} {\bibinfo
  {journal} {Rev. Mod. Phys.}\ }\textbf {\bibinfo {volume} {69}},\ \bibinfo
  {pages} {315--333} (\bibinfo {year} {1997})}\BibitemShut {NoStop}%
\bibitem [{\citenamefont {Fisher}\ and\ \citenamefont
  {Barber}(1972)}]{finiteCorr1}%
  \BibitemOpen
  \bibfield  {author} {\bibinfo {author} {\bibfnamefont {Michael~E.}\
  \bibnamefont {Fisher}}\ and\ \bibinfo {author} {\bibfnamefont {Michael~N.}\
  \bibnamefont {Barber}},\ }\bibfield  {title} {\enquote {\bibinfo {title}
  {Scaling theory for finite-size effects in the critical region},}\ }\href
  {\doibase 10.1103/PhysRevLett.28.1516} {\bibfield  {journal} {\bibinfo
  {journal} {Phys. Rev. Lett.}\ }\textbf {\bibinfo {volume} {28}},\ \bibinfo
  {pages} {1516--1519} (\bibinfo {year} {1972})}\BibitemShut {NoStop}%
\bibitem [{\citenamefont {Br{\'e}zin}\ and\ \citenamefont
  {Zinn-Justin}(1985)}]{finiteCorr2}%
  \BibitemOpen
  \bibfield  {author} {\bibinfo {author} {\bibfnamefont {E}~\bibnamefont
  {Br{\'e}zin}}\ and\ \bibinfo {author} {\bibfnamefont {Jean}\ \bibnamefont
  {Zinn-Justin}},\ }\bibfield  {title} {\enquote {\bibinfo {title} {Finite size
  effects in phase transitions},}\ }\href@noop {} {\bibfield  {journal}
  {\bibinfo  {journal} {Nuclear Physics B}\ }\textbf {\bibinfo {volume}
  {257}},\ \bibinfo {pages} {867--893} (\bibinfo {year} {1985})}\BibitemShut
  {NoStop}%
\bibitem [{\citenamefont {Osterloh}\ \emph {et~al.}(2002)\citenamefont
  {Osterloh}, \citenamefont {Amico}, \citenamefont {Falci},\ and\ \citenamefont
  {Fazio}}]{finitesize1}%
  \BibitemOpen
  \bibfield  {author} {\bibinfo {author} {\bibfnamefont {Andreas}\ \bibnamefont
  {Osterloh}}, \bibinfo {author} {\bibfnamefont {Luigi}\ \bibnamefont {Amico}},
  \bibinfo {author} {\bibfnamefont {Giuseppe}\ \bibnamefont {Falci}}, \ and\
  \bibinfo {author} {\bibfnamefont {Rosario}\ \bibnamefont {Fazio}},\
  }\bibfield  {title} {\enquote {\bibinfo {title} {Scaling of entanglement
  close to a quantum phase transition},}\ }\href@noop {} {\bibfield  {journal}
  {\bibinfo  {journal} {Nature}\ }\textbf {\bibinfo {volume} {416}},\ \bibinfo
  {pages} {608--610} (\bibinfo {year} {2002})}\BibitemShut {NoStop}%
\bibitem [{\citenamefont {Zhu}\ \emph {et~al.}(2018)\citenamefont {Zhu},
  \citenamefont {Sun}, \citenamefont {You},\ and\ \citenamefont
  {Shi}}]{finitesize2}%
  \BibitemOpen
  \bibfield  {author} {\bibinfo {author} {\bibfnamefont {Zhangqi}\ \bibnamefont
  {Zhu}}, \bibinfo {author} {\bibfnamefont {Gaoyong}\ \bibnamefont {Sun}},
  \bibinfo {author} {\bibfnamefont {Wen-Long}\ \bibnamefont {You}}, \ and\
  \bibinfo {author} {\bibfnamefont {Da-Ning}\ \bibnamefont {Shi}},\ }\bibfield
  {title} {\enquote {\bibinfo {title} {Fidelity and criticality of a quantum
  ising chain with long-range interactions},}\ }\href {\doibase
  10.1103/PhysRevA.98.023607} {\bibfield  {journal} {\bibinfo  {journal} {Phys.
  Rev. A}\ }\textbf {\bibinfo {volume} {98}},\ \bibinfo {pages} {023607}
  (\bibinfo {year} {2018})}\BibitemShut {NoStop}%
\bibitem [{\citenamefont {Hwang}\ and\ \citenamefont
  {Plenio}(2016)}]{finitesize3}%
  \BibitemOpen
  \bibfield  {author} {\bibinfo {author} {\bibfnamefont {Myung-Joong}\
  \bibnamefont {Hwang}}\ and\ \bibinfo {author} {\bibfnamefont {Martin~B.}\
  \bibnamefont {Plenio}},\ }\bibfield  {title} {\enquote {\bibinfo {title}
  {Quantum phase transition in the finite jaynes-cummings lattice systems},}\
  }\href {\doibase 10.1103/PhysRevLett.117.123602} {\bibfield  {journal}
  {\bibinfo  {journal} {Phys. Rev. Lett.}\ }\textbf {\bibinfo {volume} {117}},\
  \bibinfo {pages} {123602} (\bibinfo {year} {2016})}\BibitemShut {NoStop}%
\bibitem [{\citenamefont {Botet}\ and\ \citenamefont
  {Jullien}(1983)}]{finiteCorr3}%
  \BibitemOpen
  \bibfield  {author} {\bibinfo {author} {\bibfnamefont {R.}~\bibnamefont
  {Botet}}\ and\ \bibinfo {author} {\bibfnamefont {R.}~\bibnamefont
  {Jullien}},\ }\bibfield  {title} {\enquote {\bibinfo {title} {Large-size
  critical behavior of infinitely coordinated systems},}\ }\href {\doibase
  10.1103/PhysRevB.28.3955} {\bibfield  {journal} {\bibinfo  {journal} {Phys.
  Rev. B}\ }\textbf {\bibinfo {volume} {28}},\ \bibinfo {pages} {3955--3967}
  (\bibinfo {year} {1983})}\BibitemShut {NoStop}%
\bibitem [{\citenamefont {Cong}\ \emph {et~al.}(2019)\citenamefont {Cong},
  \citenamefont {Choi},\ and\ \citenamefont {Lukin}}]{Cong_19}%
  \BibitemOpen
  \bibfield  {author} {\bibinfo {author} {\bibfnamefont {Iris}\ \bibnamefont
  {Cong}}, \bibinfo {author} {\bibfnamefont {Soonwon}\ \bibnamefont {Choi}}, \
  and\ \bibinfo {author} {\bibfnamefont {Mikhail~D.}\ \bibnamefont {Lukin}},\
  }\bibfield  {title} {\enquote {\bibinfo {title} {Quantum convolutional neural
  networks},}\ }\href {\doibase https://doi.org/10.1038/s41567-019-0648-8}
  {\bibfield  {journal} {\bibinfo  {journal} {Nat. Phys.}\ }\textbf {\bibinfo
  {volume} {15}},\ \bibinfo {pages} {1273–1278} (\bibinfo {year}
  {2019})}\BibitemShut {NoStop}%
\bibitem [{\citenamefont {Herrmann}\ \emph {et~al.}(2022)\citenamefont
  {Herrmann}, \citenamefont {Llima}, \citenamefont {Remm}, \citenamefont
  {Zapletal}, \citenamefont {McMahon}, \citenamefont {Scarato}, \citenamefont
  {Swiadek}, \citenamefont {Andersen}, \citenamefont {Hellings}, \citenamefont
  {Krinner}, \citenamefont {Lacroix}, \citenamefont {Lazar}, \citenamefont
  {Kerschbaum}, \citenamefont {Zanuz}, \citenamefont {Norris}, \citenamefont
  {Hartmann}, \citenamefont {Wallraff},\ and\ \citenamefont
  {Eichler}}]{Wallraff_22_QPD}%
  \BibitemOpen
  \bibfield  {author} {\bibinfo {author} {\bibfnamefont {Johannes}\
  \bibnamefont {Herrmann}}, \bibinfo {author} {\bibfnamefont {Sergi~Masot}\
  \bibnamefont {Llima}}, \bibinfo {author} {\bibfnamefont {Ants}\ \bibnamefont
  {Remm}}, \bibinfo {author} {\bibfnamefont {Petr}\ \bibnamefont {Zapletal}},
  \bibinfo {author} {\bibfnamefont {Nathan~A.}\ \bibnamefont {McMahon}},
  \bibinfo {author} {\bibfnamefont {Colin}\ \bibnamefont {Scarato}}, \bibinfo
  {author} {\bibfnamefont {François}\ \bibnamefont {Swiadek}}, \bibinfo
  {author} {\bibfnamefont {Christian~Kraglund}\ \bibnamefont {Andersen}},
  \bibinfo {author} {\bibfnamefont {Christoph}\ \bibnamefont {Hellings}},
  \bibinfo {author} {\bibfnamefont {Sebastian}\ \bibnamefont {Krinner}},
  \bibinfo {author} {\bibfnamefont {Nathan}\ \bibnamefont {Lacroix}}, \bibinfo
  {author} {\bibfnamefont {Stefania}\ \bibnamefont {Lazar}}, \bibinfo {author}
  {\bibfnamefont {Michael}\ \bibnamefont {Kerschbaum}}, \bibinfo {author}
  {\bibfnamefont {Dante~Colao}\ \bibnamefont {Zanuz}}, \bibinfo {author}
  {\bibfnamefont {Graham~J.}\ \bibnamefont {Norris}}, \bibinfo {author}
  {\bibfnamefont {Michael~J.}\ \bibnamefont {Hartmann}}, \bibinfo {author}
  {\bibfnamefont {Andreas}\ \bibnamefont {Wallraff}}, \ and\ \bibinfo {author}
  {\bibfnamefont {Christopher}\ \bibnamefont {Eichler}},\ }\bibfield  {title}
  {\enquote {\bibinfo {title} {Realizing quantum convolutional neural networks
  on a superconducting quantum processor to recognize quantum phases},}\ }\href
  {\doibase https://doi.org/10.1038/s41467-022-31679-5} {\bibfield  {journal}
  {\bibinfo  {journal} {Nat Commun}\ }\textbf {\bibinfo {volume} {13}}
  (\bibinfo {year} {2022}),\
  https://doi.org/10.1038/s41467-022-31679-5}\BibitemShut {NoStop}%
\bibitem [{\citenamefont {Kottmann}\ \emph {et~al.}(2021)\citenamefont
  {Kottmann}, \citenamefont {Metz}, \citenamefont {Fraxanet},\ and\
  \citenamefont {Baldelli}}]{PRR_anomaly_detection}%
  \BibitemOpen
  \bibfield  {author} {\bibinfo {author} {\bibfnamefont {Korbinian}\
  \bibnamefont {Kottmann}}, \bibinfo {author} {\bibfnamefont {Friederike}\
  \bibnamefont {Metz}}, \bibinfo {author} {\bibfnamefont {Joana}\ \bibnamefont
  {Fraxanet}}, \ and\ \bibinfo {author} {\bibfnamefont {Niccol\`o}\
  \bibnamefont {Baldelli}},\ }\bibfield  {title} {\enquote {\bibinfo {title}
  {Variational quantum anomaly detection: Unsupervised mapping of phase
  diagrams on a physical quantum computer},}\ }\href {\doibase
  10.1103/PhysRevResearch.3.043184} {\bibfield  {journal} {\bibinfo  {journal}
  {Phys. Rev. Research}\ }\textbf {\bibinfo {volume} {3}},\ \bibinfo {pages}
  {043184} (\bibinfo {year} {2021})}\BibitemShut {NoStop}%
\bibitem [{\citenamefont {Gibbs}\ \emph {et~al.}(2022)\citenamefont {Gibbs},
  \citenamefont {Holmes}, \citenamefont {Caro}, \citenamefont {Ezzell},
  \citenamefont {Huang}, \citenamefont {Cincio}, \citenamefont {Sornborger},\
  and\ \citenamefont {Coles}}]{caroQML}%
  \BibitemOpen
  \bibfield  {author} {\bibinfo {author} {\bibfnamefont {Joe}\ \bibnamefont
  {Gibbs}}, \bibinfo {author} {\bibfnamefont {Zoe}\ \bibnamefont {Holmes}},
  \bibinfo {author} {\bibfnamefont {Matthias~C.}\ \bibnamefont {Caro}},
  \bibinfo {author} {\bibfnamefont {Nicholas}\ \bibnamefont {Ezzell}}, \bibinfo
  {author} {\bibfnamefont {Hsin-Yuan}\ \bibnamefont {Huang}}, \bibinfo {author}
  {\bibfnamefont {Lukasz}\ \bibnamefont {Cincio}}, \bibinfo {author}
  {\bibfnamefont {Andrew~T.}\ \bibnamefont {Sornborger}}, \ and\ \bibinfo
  {author} {\bibfnamefont {Patrick~J.}\ \bibnamefont {Coles}},\ }\bibfield
  {title} {\enquote {\bibinfo {title} {Dynamical simulation via quantum machine
  learning with provable generalization},}\ }\href {\doibase
  10.48550/ARXIV.2204.10269} {\  (\bibinfo {year} {2022}),\
  10.48550/ARXIV.2204.10269}\BibitemShut {NoStop}%
\bibitem [{\citenamefont {Daley}\ \emph {et~al.}(2022)\citenamefont {Daley},
  \citenamefont {Bloch}, \citenamefont {Kokail}, \citenamefont {Flannigan},
  \citenamefont {Pearson}, \citenamefont {Troyer},\ and\ \citenamefont
  {Zoller}}]{zollerqadv}%
  \BibitemOpen
  \bibfield  {author} {\bibinfo {author} {\bibfnamefont {Andrew}\ \bibnamefont
  {Daley}}, \bibinfo {author} {\bibfnamefont {Immanuel}\ \bibnamefont {Bloch}},
  \bibinfo {author} {\bibfnamefont {Christian}\ \bibnamefont {Kokail}},
  \bibinfo {author} {\bibfnamefont {Stuart}\ \bibnamefont {Flannigan}},
  \bibinfo {author} {\bibfnamefont {Natalie}\ \bibnamefont {Pearson}}, \bibinfo
  {author} {\bibfnamefont {Matthias}\ \bibnamefont {Troyer}}, \ and\ \bibinfo
  {author} {\bibfnamefont {Peter}\ \bibnamefont {Zoller}},\ }\bibfield  {title}
  {\enquote {\bibinfo {title} {Practical quantum advantage in quantum
  simulation},}\ }\href {\doibase 10.1038/s41586-022-04940-6} {\bibfield
  {journal} {\bibinfo  {journal} {Nature}\ }\textbf {\bibinfo {volume} {607}},\
  \bibinfo {pages} {667--676} (\bibinfo {year} {2022})}\BibitemShut {NoStop}%
\bibitem [{\citenamefont {{Lipkin}}\ \emph {et~al.}(1965)\citenamefont
  {{Lipkin}}, \citenamefont {{Meshkov}},\ and\ \citenamefont {{Glick}}}]{LMG1}%
  \BibitemOpen
  \bibfield  {author} {\bibinfo {author} {\bibfnamefont {Harry~J}\ \bibnamefont
  {{Lipkin}}}, \bibinfo {author} {\bibfnamefont {N}~\bibnamefont {{Meshkov}}},
  \ and\ \bibinfo {author} {\bibfnamefont {AJ}~\bibnamefont {{Glick}}},\
  }\bibfield  {title} {\enquote {\bibinfo {title} {Validity of many-body
  approximation methods for a solvable model:(i). exact solutions and
  perturbation theory},}\ }\href@noop {} {\bibfield  {journal} {\bibinfo
  {journal} {Nuclear Physics}\ }\textbf {\bibinfo {volume} {62}},\ \bibinfo
  {pages} {188--198} (\bibinfo {year} {1965})}\BibitemShut {NoStop}%
\bibitem [{\citenamefont {{Meshkov}}\ \emph {et~al.}(1965)\citenamefont
  {{Meshkov}}, \citenamefont {{Glick}},\ and\ \citenamefont {{Lipkin}}}]{LMG2}%
  \BibitemOpen
  \bibfield  {author} {\bibinfo {author} {\bibfnamefont {N}~\bibnamefont
  {{Meshkov}}}, \bibinfo {author} {\bibfnamefont {AJ}~\bibnamefont {{Glick}}},
  \ and\ \bibinfo {author} {\bibfnamefont {HJ}~\bibnamefont {{Lipkin}}},\
  }\bibfield  {title} {\enquote {\bibinfo {title} {Validity of many-body
  approximation methods for a solvable model:(ii). linearization procedures},}\
  }\href@noop {} {\bibfield  {journal} {\bibinfo  {journal} {Nuclear Physics}\
  }\textbf {\bibinfo {volume} {62}},\ \bibinfo {pages} {199--210} (\bibinfo
  {year} {1965})}\BibitemShut {NoStop}%
\bibitem [{\citenamefont {{Glick}}\ \emph {et~al.}(1965)\citenamefont
  {{Glick}}, \citenamefont {{Lipkin}},\ and\ \citenamefont {{Meshkov}}}]{LMG3}%
  \BibitemOpen
  \bibfield  {author} {\bibinfo {author} {\bibfnamefont {AJ}~\bibnamefont
  {{Glick}}}, \bibinfo {author} {\bibfnamefont {HJ}~\bibnamefont {{Lipkin}}}, \
  and\ \bibinfo {author} {\bibfnamefont {N}~\bibnamefont {{Meshkov}}},\
  }\bibfield  {title} {\enquote {\bibinfo {title} {Validity of many-body
  approximation methods for a solvable model:(iii). diagram summations},}\
  }\href@noop {} {\bibfield  {journal} {\bibinfo  {journal} {Nuclear Physics}\
  }\textbf {\bibinfo {volume} {62}},\ \bibinfo {pages} {211--224} (\bibinfo
  {year} {1965})}\BibitemShut {NoStop}%
\bibitem [{\citenamefont {Preskill}(2018)}]{Preskill2018quantumcomputingin}%
  \BibitemOpen
  \bibfield  {author} {\bibinfo {author} {\bibfnamefont {John}\ \bibnamefont
  {Preskill}},\ }\bibfield  {title} {\enquote {\bibinfo {title} {Quantum
  {C}omputing in the {NISQ} era and beyond},}\ }\href {\doibase
  10.22331/q-2018-08-06-79} {\bibfield  {journal} {\bibinfo  {journal}
  {{Quantum}}\ }\textbf {\bibinfo {volume} {2}},\ \bibinfo {pages} {79}
  (\bibinfo {year} {2018})}\BibitemShut {NoStop}%
\bibitem [{\citenamefont {Bharti}\ \emph {et~al.}(2022)\citenamefont {Bharti},
  \citenamefont {Cervera-Lierta}, \citenamefont {Kyaw}, \citenamefont {Haug},
  \citenamefont {Alperin-Lea}, \citenamefont {Anand}, \citenamefont {Degroote},
  \citenamefont {Heimonen}, \citenamefont {Kottmann}, \citenamefont {Menke},
  \citenamefont {Mok}, \citenamefont {Sim}, \citenamefont {Kwek},\ and\
  \citenamefont {Aspuru-Guzik}}]{RMP_NISQ}%
  \BibitemOpen
  \bibfield  {author} {\bibinfo {author} {\bibfnamefont {Kishor}\ \bibnamefont
  {Bharti}}, \bibinfo {author} {\bibfnamefont {Alba}\ \bibnamefont
  {Cervera-Lierta}}, \bibinfo {author} {\bibfnamefont {Thi~Ha}\ \bibnamefont
  {Kyaw}}, \bibinfo {author} {\bibfnamefont {Tobias}\ \bibnamefont {Haug}},
  \bibinfo {author} {\bibfnamefont {Sumner}\ \bibnamefont {Alperin-Lea}},
  \bibinfo {author} {\bibfnamefont {Abhinav}\ \bibnamefont {Anand}}, \bibinfo
  {author} {\bibfnamefont {Matthias}\ \bibnamefont {Degroote}}, \bibinfo
  {author} {\bibfnamefont {Hermanni}\ \bibnamefont {Heimonen}}, \bibinfo
  {author} {\bibfnamefont {Jakob~S.}\ \bibnamefont {Kottmann}}, \bibinfo
  {author} {\bibfnamefont {Tim}\ \bibnamefont {Menke}}, \bibinfo {author}
  {\bibfnamefont {Wai-Keong}\ \bibnamefont {Mok}}, \bibinfo {author}
  {\bibfnamefont {Sukin}\ \bibnamefont {Sim}}, \bibinfo {author} {\bibfnamefont
  {Leong-Chuan}\ \bibnamefont {Kwek}}, \ and\ \bibinfo {author} {\bibfnamefont
  {Al\'an}\ \bibnamefont {Aspuru-Guzik}},\ }\bibfield  {title} {\enquote
  {\bibinfo {title} {Noisy intermediate-scale quantum algorithms},}\ }\href
  {\doibase 10.1103/RevModPhys.94.015004} {\bibfield  {journal} {\bibinfo
  {journal} {Rev. Mod. Phys.}\ }\textbf {\bibinfo {volume} {94}},\ \bibinfo
  {pages} {015004} (\bibinfo {year} {2022})}\BibitemShut {NoStop}%
\bibitem [{\citenamefont {Uvarov}\ \emph
  {et~al.}(2020{\natexlab{a}})\citenamefont {Uvarov}, \citenamefont
  {Biamonte},\ and\ \citenamefont {Yudin}}]{VQE_QPT}%
  \BibitemOpen
  \bibfield  {author} {\bibinfo {author} {\bibfnamefont {Alexey}\ \bibnamefont
  {Uvarov}}, \bibinfo {author} {\bibfnamefont {Jacob~D.}\ \bibnamefont
  {Biamonte}}, \ and\ \bibinfo {author} {\bibfnamefont {Dmitry}\ \bibnamefont
  {Yudin}},\ }\bibfield  {title} {\enquote {\bibinfo {title} {Variational
  quantum eigensolver for frustrated quantum systems},}\ }\href {\doibase
  10.1103/PhysRevB.102.075104} {\bibfield  {journal} {\bibinfo  {journal}
  {Phys. Rev. B}\ }\textbf {\bibinfo {volume} {102}},\ \bibinfo {pages}
  {075104} (\bibinfo {year} {2020}{\natexlab{a}})}\BibitemShut {NoStop}%
\bibitem [{\citenamefont {Casta\~nos}\ \emph {et~al.}(2005)\citenamefont
  {Casta\~nos}, \citenamefont {L\'opez-Pe\~na}, \citenamefont {Hirsch},\ and\
  \citenamefont {L\'opez-Moreno}}]{LMG_nano}%
  \BibitemOpen
  \bibfield  {author} {\bibinfo {author} {\bibfnamefont {Octavio}\ \bibnamefont
  {Casta\~nos}}, \bibinfo {author} {\bibfnamefont {Ram\'on}\ \bibnamefont
  {L\'opez-Pe\~na}}, \bibinfo {author} {\bibfnamefont {Jorge~G.}\ \bibnamefont
  {Hirsch}}, \ and\ \bibinfo {author} {\bibfnamefont {Enrique}\ \bibnamefont
  {L\'opez-Moreno}},\ }\bibfield  {title} {\enquote {\bibinfo {title} {Phase
  transitions and accidental degeneracy in nonlinear spin systems},}\ }\href
  {\doibase 10.1103/PhysRevB.72.012406} {\bibfield  {journal} {\bibinfo
  {journal} {Phys. Rev. B}\ }\textbf {\bibinfo {volume} {72}},\ \bibinfo
  {pages} {012406} (\bibinfo {year} {2005})}\BibitemShut {NoStop}%
\bibitem [{\citenamefont {Kwok}\ \emph {et~al.}(2008)\citenamefont {Kwok},
  \citenamefont {Ning}, \citenamefont {Gu},\ and\ \citenamefont
  {Lin}}]{LMG_susc1}%
  \BibitemOpen
  \bibfield  {author} {\bibinfo {author} {\bibfnamefont {Ho-Man}\ \bibnamefont
  {Kwok}}, \bibinfo {author} {\bibfnamefont {Wen-Qiang}\ \bibnamefont {Ning}},
  \bibinfo {author} {\bibfnamefont {Shi-Jian}\ \bibnamefont {Gu}}, \ and\
  \bibinfo {author} {\bibfnamefont {Hai-Qing}\ \bibnamefont {Lin}},\ }\bibfield
   {title} {\enquote {\bibinfo {title} {Quantum criticality of the
  {Lipkin}-{Meshkov}-{Glick} model in terms of fidelity susceptibility},}\
  }\href {\doibase 10.1103/PhysRevE.78.032103} {\bibfield  {journal} {\bibinfo
  {journal} {Phys. Rev. E}\ }\textbf {\bibinfo {volume} {78}},\ \bibinfo
  {pages} {032103} (\bibinfo {year} {2008})}\BibitemShut {NoStop}%
\bibitem [{\citenamefont {Ma}\ \emph {et~al.}(2009)\citenamefont {Ma},
  \citenamefont {Wang},\ and\ \citenamefont {Gu}}]{LMG_susc2}%
  \BibitemOpen
  \bibfield  {author} {\bibinfo {author} {\bibfnamefont {Jian}\ \bibnamefont
  {Ma}}, \bibinfo {author} {\bibfnamefont {Xiaoguang}\ \bibnamefont {Wang}}, \
  and\ \bibinfo {author} {\bibfnamefont {Shi-Jian}\ \bibnamefont {Gu}},\
  }\bibfield  {title} {\enquote {\bibinfo {title} {Many-body reduced fidelity
  susceptibility in {Lipkin}-{Meshkov}-{Glick} model},}\ }\href {\doibase
  10.1103/PhysRevE.80.021124} {\bibfield  {journal} {\bibinfo  {journal} {Phys.
  Rev. E}\ }\textbf {\bibinfo {volume} {80}},\ \bibinfo {pages} {021124}
  (\bibinfo {year} {2009})}\BibitemShut {NoStop}%
\bibitem [{\citenamefont {Gonzalez}\ \emph {et~al.}(2021)\citenamefont
  {Gonzalez}, \citenamefont {Guti\'errez-Ruiz},\ and\ \citenamefont
  {Vergara}}]{LMG_geom}%
  \BibitemOpen
  \bibfield  {author} {\bibinfo {author} {\bibfnamefont {Diego}\ \bibnamefont
  {Gonzalez}}, \bibinfo {author} {\bibfnamefont {Daniel}\ \bibnamefont
  {Guti\'errez-Ruiz}}, \ and\ \bibinfo {author} {\bibfnamefont {J.~David}\
  \bibnamefont {Vergara}},\ }\bibfield  {title} {\enquote {\bibinfo {title}
  {Classical description of the parameter space geometry in the {Dicke} and
  {Lipkin}-{Meshkov}-{Glick} models},}\ }\href {\doibase
  10.1103/PhysRevE.104.014113} {\bibfield  {journal} {\bibinfo  {journal}
  {Phys. Rev. E}\ }\textbf {\bibinfo {volume} {104}},\ \bibinfo {pages}
  {014113} (\bibinfo {year} {2021})}\BibitemShut {NoStop}%
\bibitem [{\citenamefont {Cervia}\ \emph {et~al.}(2021)\citenamefont {Cervia},
  \citenamefont {Balantekin}, \citenamefont {Coppersmith}, \citenamefont
  {Johnson}, \citenamefont {Love}, \citenamefont {Poole}, \citenamefont
  {Robbins},\ and\ \citenamefont {Saffman}}]{LMG_QC1}%
  \BibitemOpen
  \bibfield  {author} {\bibinfo {author} {\bibfnamefont {Michael~J.}\
  \bibnamefont {Cervia}}, \bibinfo {author} {\bibfnamefont {A.~B.}\
  \bibnamefont {Balantekin}}, \bibinfo {author} {\bibfnamefont {S.~N.}\
  \bibnamefont {Coppersmith}}, \bibinfo {author} {\bibfnamefont {Calvin~W.}\
  \bibnamefont {Johnson}}, \bibinfo {author} {\bibfnamefont {Peter~J.}\
  \bibnamefont {Love}}, \bibinfo {author} {\bibfnamefont {C.}~\bibnamefont
  {Poole}}, \bibinfo {author} {\bibfnamefont {K.}~\bibnamefont {Robbins}}, \
  and\ \bibinfo {author} {\bibfnamefont {M.}~\bibnamefont {Saffman}},\
  }\bibfield  {title} {\enquote {\bibinfo {title} {{Lipkin} model on a quantum
  computer},}\ }\href {\doibase 10.1103/PhysRevC.104.024305} {\bibfield
  {journal} {\bibinfo  {journal} {Phys. Rev. C}\ }\textbf {\bibinfo {volume}
  {104}},\ \bibinfo {pages} {024305} (\bibinfo {year} {2021})}\BibitemShut
  {NoStop}%
\bibitem [{\citenamefont {Robbins}\ and\ \citenamefont {Love}(2021)}]{LMG_QC2}%
  \BibitemOpen
  \bibfield  {author} {\bibinfo {author} {\bibfnamefont {Kenneth}\ \bibnamefont
  {Robbins}}\ and\ \bibinfo {author} {\bibfnamefont {Peter~J.}\ \bibnamefont
  {Love}},\ }\bibfield  {title} {\enquote {\bibinfo {title} {Benchmarking
  near-term quantum devices with the variational quantum eigensolver and the
  {Lipkin}-{Meshkov}-{Glick} model},}\ }\href {\doibase
  10.1103/PhysRevA.104.022412} {\bibfield  {journal} {\bibinfo  {journal}
  {Phys. Rev. A}\ }\textbf {\bibinfo {volume} {104}},\ \bibinfo {pages}
  {022412} (\bibinfo {year} {2021})}\BibitemShut {NoStop}%
\bibitem [{\citenamefont {Chikaoka}\ and\ \citenamefont
  {Liang}(2022)}]{LMG_QC3}%
  \BibitemOpen
  \bibfield  {author} {\bibinfo {author} {\bibfnamefont {Asahi}\ \bibnamefont
  {Chikaoka}}\ and\ \bibinfo {author} {\bibfnamefont {Haozhao}\ \bibnamefont
  {Liang}},\ }\bibfield  {title} {\enquote {\bibinfo {title} {Quantum computing
  for the {Lipkin} model with unitary coupled cluster and structure learning
  ansatz},}\ }\href {\doibase 10.1088/1674-1137/ac380a} {\bibfield  {journal}
  {\bibinfo  {journal} {Chinese Physics C}\ }\textbf {\bibinfo {volume} {46}},\
  \bibinfo {pages} {024106} (\bibinfo {year} {2022})}\BibitemShut {NoStop}%
\bibitem [{\citenamefont {Romero}\ \emph {et~al.}(2022)\citenamefont {Romero},
  \citenamefont {Engel}, \citenamefont {Tang},\ and\ \citenamefont
  {Economou}}]{LMG_QC4}%
  \BibitemOpen
  \bibfield  {author} {\bibinfo {author} {\bibfnamefont {A.~M.}\ \bibnamefont
  {Romero}}, \bibinfo {author} {\bibfnamefont {J.}~\bibnamefont {Engel}},
  \bibinfo {author} {\bibfnamefont {Ho~Lun}\ \bibnamefont {Tang}}, \ and\
  \bibinfo {author} {\bibfnamefont {Sophia~E.}\ \bibnamefont {Economou}},\
  }\bibfield  {title} {\enquote {\bibinfo {title} {Solving nuclear structure
  problems with the adaptive variational quantum algorithm},}\ }\href {\doibase
  10.1103/PhysRevC.105.064317} {\bibfield  {journal} {\bibinfo  {journal}
  {Phys. Rev. C}\ }\textbf {\bibinfo {volume} {105}},\ \bibinfo {pages}
  {064317} (\bibinfo {year} {2022})}\BibitemShut {NoStop}%
\bibitem [{\citenamefont {Peruzzo}\ \emph {et~al.}(2014)\citenamefont
  {Peruzzo}, \citenamefont {McClean}, \citenamefont {Shadbolt}, \citenamefont
  {Yung}, \citenamefont {Zhou}, \citenamefont {Love}, \citenamefont
  {Aspuru-Guzik},\ and\ \citenamefont {O’Brien}}]{vqe_original}%
  \BibitemOpen
  \bibfield  {author} {\bibinfo {author} {\bibfnamefont {Alberto}\ \bibnamefont
  {Peruzzo}}, \bibinfo {author} {\bibfnamefont {Jarrod}\ \bibnamefont
  {McClean}}, \bibinfo {author} {\bibfnamefont {Peter}\ \bibnamefont
  {Shadbolt}}, \bibinfo {author} {\bibfnamefont {Man-Hong}\ \bibnamefont
  {Yung}}, \bibinfo {author} {\bibfnamefont {Xiao-Qi}\ \bibnamefont {Zhou}},
  \bibinfo {author} {\bibfnamefont {Peter~J.}\ \bibnamefont {Love}}, \bibinfo
  {author} {\bibfnamefont {Al\'an}\ \bibnamefont {Aspuru-Guzik}}, \ and\
  \bibinfo {author} {\bibfnamefont {Jeremy~L.}\ \bibnamefont {O’Brien}},\
  }\bibfield  {title} {\enquote {\bibinfo {title} {A variational eigenvalue
  solver on a photonic quantum processor},}\ }\href {\doibase
  https://doi.org/10.1038/ncomms5213} {\bibfield  {journal} {\bibinfo
  {journal} {Nature Communications}\ }\textbf {\bibinfo {volume} {5}},\
  \bibinfo {pages} {4123} (\bibinfo {year} {2014})}\BibitemShut {NoStop}%
\bibitem [{\citenamefont {Tilly}\ \emph {et~al.}(2022)\citenamefont {Tilly},
  \citenamefont {Chen}, \citenamefont {Cao}, \citenamefont {Picozzi},
  \citenamefont {Setia}, \citenamefont {Li}, \citenamefont {Grant},
  \citenamefont {Wossnig}, \citenamefont {Rungger}, \citenamefont {Booth} \emph
  {et~al.}}]{Rev_VQE}%
  \BibitemOpen
  \bibfield  {author} {\bibinfo {author} {\bibfnamefont {Jules}\ \bibnamefont
  {Tilly}}, \bibinfo {author} {\bibfnamefont {Hongxiang}\ \bibnamefont {Chen}},
  \bibinfo {author} {\bibfnamefont {Shuxiang}\ \bibnamefont {Cao}}, \bibinfo
  {author} {\bibfnamefont {Dario}\ \bibnamefont {Picozzi}}, \bibinfo {author}
  {\bibfnamefont {Kanav}\ \bibnamefont {Setia}}, \bibinfo {author}
  {\bibfnamefont {Ying}\ \bibnamefont {Li}}, \bibinfo {author} {\bibfnamefont
  {Edward}\ \bibnamefont {Grant}}, \bibinfo {author} {\bibfnamefont {Leonard}\
  \bibnamefont {Wossnig}}, \bibinfo {author} {\bibfnamefont {Ivan}\
  \bibnamefont {Rungger}}, \bibinfo {author} {\bibfnamefont {George~H}\
  \bibnamefont {Booth}},  \emph {et~al.},\ }\bibfield  {title} {\enquote
  {\bibinfo {title} {The variational quantum eigensolver: a review of methods
  and best practices},}\ }\href@noop {} {\bibfield  {journal} {\bibinfo
  {journal} {Physics Reports}\ }\textbf {\bibinfo {volume} {986}},\ \bibinfo
  {pages} {1--128} (\bibinfo {year} {2022})}\BibitemShut {NoStop}%
\bibitem [{\citenamefont {McClean}\ \emph {et~al.}(2016)\citenamefont
  {McClean}, \citenamefont {Romero}, \citenamefont {Babbush},\ and\
  \citenamefont {Aspuru-Guzik}}]{variational_quantum_algorithm}%
  \BibitemOpen
  \bibfield  {author} {\bibinfo {author} {\bibfnamefont {Jarrod~R}\
  \bibnamefont {McClean}}, \bibinfo {author} {\bibfnamefont {Jonathan}\
  \bibnamefont {Romero}}, \bibinfo {author} {\bibfnamefont {Ryan}\ \bibnamefont
  {Babbush}}, \ and\ \bibinfo {author} {\bibfnamefont {Al{\'{a}}n}\
  \bibnamefont {Aspuru-Guzik}},\ }\bibfield  {title} {\enquote {\bibinfo
  {title} {The theory of variational hybrid quantum-classical algorithms},}\
  }\href {\doibase 10.1088/1367-2630/18/2/023023} {\bibfield  {journal}
  {\bibinfo  {journal} {New Journal of Physics}\ }\textbf {\bibinfo {volume}
  {18}},\ \bibinfo {pages} {023023} (\bibinfo {year} {2016})}\BibitemShut
  {NoStop}%
\bibitem [{\citenamefont {Romero}\ \emph {et~al.}(2019)\citenamefont {Romero},
  \citenamefont {Babbush}, \citenamefont {McClean}, \citenamefont {Hempel},
  \citenamefont {Love},\ and\ \citenamefont {Aspuru-Guzik}}]{UCC-chemistry}%
  \BibitemOpen
  \bibfield  {author} {\bibinfo {author} {\bibfnamefont {Jonathan}\
  \bibnamefont {Romero}}, \bibinfo {author} {\bibfnamefont {Ryan}\ \bibnamefont
  {Babbush}}, \bibinfo {author} {\bibfnamefont {Jarrod~R}\ \bibnamefont
  {McClean}}, \bibinfo {author} {\bibfnamefont {Cornelius}\ \bibnamefont
  {Hempel}}, \bibinfo {author} {\bibfnamefont {Peter~J}\ \bibnamefont {Love}},
  \ and\ \bibinfo {author} {\bibfnamefont {Al\'an}\ \bibnamefont
  {Aspuru-Guzik}},\ }\bibfield  {title} {\enquote {\bibinfo {title} {Strategies
  for quantum computing molecular energies using the unitary coupled cluster
  ansatz},}\ }\href {\doibase 10.1088/2058-9565/aad3e4} {\bibfield  {journal}
  {\bibinfo  {journal} {Quantum Sci. Technol.}\ }\textbf {\bibinfo {volume}
  {4}},\ \bibinfo {pages} {014008} (\bibinfo {year} {2019})}\BibitemShut
  {NoStop}%
\bibitem [{\citenamefont {Kandala}\ \emph {et~al.}(2017)\citenamefont
  {Kandala}, \citenamefont {Mezzacapo}, \citenamefont {Temme}, \citenamefont
  {Takita}, \citenamefont {Brink}, \citenamefont {Chow},\ and\ \citenamefont
  {Gambetta}}]{VQE_Gambetta}%
  \BibitemOpen
  \bibfield  {author} {\bibinfo {author} {\bibfnamefont {Abhinav}\ \bibnamefont
  {Kandala}}, \bibinfo {author} {\bibfnamefont {Antonio}\ \bibnamefont
  {Mezzacapo}}, \bibinfo {author} {\bibfnamefont {Kristan}\ \bibnamefont
  {Temme}}, \bibinfo {author} {\bibfnamefont {Maika}\ \bibnamefont {Takita}},
  \bibinfo {author} {\bibfnamefont {Markus}\ \bibnamefont {Brink}}, \bibinfo
  {author} {\bibfnamefont {Jerry~M.}\ \bibnamefont {Chow}}, \ and\ \bibinfo
  {author} {\bibfnamefont {Jay~M.}\ \bibnamefont {Gambetta}},\ }\bibfield
  {title} {\enquote {\bibinfo {title} {Hardware-efficient variational quantum
  eigensolver for small molecules and quantum magnets},}\ }\href {\doibase
  https://doi.org/10.1038/nature23879} {\bibfield  {journal} {\bibinfo
  {journal} {Nature}\ }\textbf {\bibinfo {volume} {549}},\ \bibinfo {pages}
  {242--246} (\bibinfo {year} {2017})}\BibitemShut {NoStop}%
\bibitem [{\citenamefont {Crippa}\ \emph {et~al.}(2021)\citenamefont {Crippa},
  \citenamefont {Tacchino}, \citenamefont {Chizzini}, \citenamefont {Aita},
  \citenamefont {Grossi}, \citenamefont {Chiesa}, \citenamefont {Santini},
  \citenamefont {Tavernelli},\ and\ \citenamefont {Carretta}}]{magnetogrossi}%
  \BibitemOpen
  \bibfield  {author} {\bibinfo {author} {\bibfnamefont {Luca}\ \bibnamefont
  {Crippa}}, \bibinfo {author} {\bibfnamefont {Francesco}\ \bibnamefont
  {Tacchino}}, \bibinfo {author} {\bibfnamefont {Mario}\ \bibnamefont
  {Chizzini}}, \bibinfo {author} {\bibfnamefont {Antonello}\ \bibnamefont
  {Aita}}, \bibinfo {author} {\bibfnamefont {Michele}\ \bibnamefont {Grossi}},
  \bibinfo {author} {\bibfnamefont {Alessandro}\ \bibnamefont {Chiesa}},
  \bibinfo {author} {\bibfnamefont {Paolo}\ \bibnamefont {Santini}}, \bibinfo
  {author} {\bibfnamefont {Ivano}\ \bibnamefont {Tavernelli}}, \ and\ \bibinfo
  {author} {\bibfnamefont {Stefano}\ \bibnamefont {Carretta}},\ }\bibfield
  {title} {\enquote {\bibinfo {title} {Simulating static and dynamic properties
  of magnetic molecules with prototype quantum computers},}\ }\href {\doibase
  10.3390/magnetochemistry7080117} {\bibfield  {journal} {\bibinfo  {journal}
  {Magnetochemistry}\ }\textbf {\bibinfo {volume} {7}} (\bibinfo {year}
  {2021}),\ 10.3390/magnetochemistry7080117}\BibitemShut {NoStop}%
\bibitem [{\citenamefont {Barkoutsos}\ \emph {et~al.}(2018)\citenamefont
  {Barkoutsos}, \citenamefont {Gonthier}, \citenamefont {Sokolov},
  \citenamefont {Moll}, \citenamefont {Salis}, \citenamefont {Fuhrer},
  \citenamefont {Ganzhorn}, \citenamefont {Egger}, \citenamefont {Troyer},
  \citenamefont {Mezzacapo}, \citenamefont {Filipp},\ and\ \citenamefont
  {Tavernelli}}]{Panos_excitation_preserving}%
  \BibitemOpen
  \bibfield  {author} {\bibinfo {author} {\bibfnamefont {Panagiotis~Kl.}\
  \bibnamefont {Barkoutsos}}, \bibinfo {author} {\bibfnamefont {Jerome~F.}\
  \bibnamefont {Gonthier}}, \bibinfo {author} {\bibfnamefont {Igor}\
  \bibnamefont {Sokolov}}, \bibinfo {author} {\bibfnamefont {Nikolaj}\
  \bibnamefont {Moll}}, \bibinfo {author} {\bibfnamefont {Gian}\ \bibnamefont
  {Salis}}, \bibinfo {author} {\bibfnamefont {Andreas}\ \bibnamefont {Fuhrer}},
  \bibinfo {author} {\bibfnamefont {Marc}\ \bibnamefont {Ganzhorn}}, \bibinfo
  {author} {\bibfnamefont {Daniel~J.}\ \bibnamefont {Egger}}, \bibinfo {author}
  {\bibfnamefont {Matthias}\ \bibnamefont {Troyer}}, \bibinfo {author}
  {\bibfnamefont {Antonio}\ \bibnamefont {Mezzacapo}}, \bibinfo {author}
  {\bibfnamefont {Stefan}\ \bibnamefont {Filipp}}, \ and\ \bibinfo {author}
  {\bibfnamefont {Ivano}\ \bibnamefont {Tavernelli}},\ }\bibfield  {title}
  {\enquote {\bibinfo {title} {Quantum algorithms for electronic structure
  calculations: Particle-hole hamiltonian and optimized wave-function
  expansions},}\ }\href {\doibase 10.1103/PhysRevA.98.022322} {\bibfield
  {journal} {\bibinfo  {journal} {Phys. Rev. A}\ }\textbf {\bibinfo {volume}
  {98}},\ \bibinfo {pages} {022322} (\bibinfo {year} {2018})}\BibitemShut
  {NoStop}%
\bibitem [{\citenamefont {Dumitrescu}\ \emph {et~al.}(2018)\citenamefont
  {Dumitrescu}, \citenamefont {McCaskey}, \citenamefont {Hagen}, \citenamefont
  {Jansen}, \citenamefont {Morris}, \citenamefont {Papenbrock}, \citenamefont
  {Pooser}, \citenamefont {Dean},\ and\ \citenamefont
  {Lougovski}}]{Papenbrock-Deuterium}%
  \BibitemOpen
  \bibfield  {author} {\bibinfo {author} {\bibfnamefont {E.~F.}\ \bibnamefont
  {Dumitrescu}}, \bibinfo {author} {\bibfnamefont {A.~J.}\ \bibnamefont
  {McCaskey}}, \bibinfo {author} {\bibfnamefont {G.}~\bibnamefont {Hagen}},
  \bibinfo {author} {\bibfnamefont {G.~R.}\ \bibnamefont {Jansen}}, \bibinfo
  {author} {\bibfnamefont {T.~D.}\ \bibnamefont {Morris}}, \bibinfo {author}
  {\bibfnamefont {T.}~\bibnamefont {Papenbrock}}, \bibinfo {author}
  {\bibfnamefont {R.~C.}\ \bibnamefont {Pooser}}, \bibinfo {author}
  {\bibfnamefont {D.~J.}\ \bibnamefont {Dean}}, \ and\ \bibinfo {author}
  {\bibfnamefont {P.}~\bibnamefont {Lougovski}},\ }\bibfield  {title} {\enquote
  {\bibinfo {title} {Cloud quantum computing of an atomic nucleus},}\ }\href
  {\doibase 10.1103/PhysRevLett.120.210501} {\bibfield  {journal} {\bibinfo
  {journal} {Phys. Rev. Lett.}\ }\textbf {\bibinfo {volume} {120}},\ \bibinfo
  {pages} {210501} (\bibinfo {year} {2018})}\BibitemShut {NoStop}%
\bibitem [{\citenamefont {Stetcu}\ \emph {et~al.}(2022)\citenamefont {Stetcu},
  \citenamefont {Baroni},\ and\ \citenamefont {Carlson}}]{Be8-VQE}%
  \BibitemOpen
  \bibfield  {author} {\bibinfo {author} {\bibfnamefont {I.}~\bibnamefont
  {Stetcu}}, \bibinfo {author} {\bibfnamefont {A.}~\bibnamefont {Baroni}}, \
  and\ \bibinfo {author} {\bibfnamefont {J.}~\bibnamefont {Carlson}},\
  }\bibfield  {title} {\enquote {\bibinfo {title} {Variational approaches to
  constructing the many-body nuclear ground state for quantum computing},}\
  }\href {\doibase 10.1103/PhysRevC.105.064308} {\bibfield  {journal} {\bibinfo
   {journal} {Phys. Rev. C}\ }\textbf {\bibinfo {volume} {105}},\ \bibinfo
  {pages} {064308} (\bibinfo {year} {2022})}\BibitemShut {NoStop}%
\bibitem [{\citenamefont {Kiss}\ \emph {et~al.}(2022)\citenamefont {Kiss},
  \citenamefont {Grossi}, \citenamefont {Lougovski}, \citenamefont {Sanchez},
  \citenamefont {Vallecorsa},\ and\ \citenamefont {Papenbrock}}]{VQE_Li6}%
  \BibitemOpen
  \bibfield  {author} {\bibinfo {author} {\bibfnamefont {Oriel}\ \bibnamefont
  {Kiss}}, \bibinfo {author} {\bibfnamefont {Michele}\ \bibnamefont {Grossi}},
  \bibinfo {author} {\bibfnamefont {Pavel}\ \bibnamefont {Lougovski}}, \bibinfo
  {author} {\bibfnamefont {Federico}\ \bibnamefont {Sanchez}}, \bibinfo
  {author} {\bibfnamefont {Sofia}\ \bibnamefont {Vallecorsa}}, \ and\ \bibinfo
  {author} {\bibfnamefont {Thomas}\ \bibnamefont {Papenbrock}},\ }\bibfield
  {title} {\enquote {\bibinfo {title} {Quantum computing of the
  $^{6}\mathrm{Li}$ nucleus via ordered unitary coupled clusters},}\ }\href
  {\doibase 10.1103/PhysRevC.106.034325} {\bibfield  {journal} {\bibinfo
  {journal} {Phys. Rev. C}\ }\textbf {\bibinfo {volume} {106}},\ \bibinfo
  {pages} {034325} (\bibinfo {year} {2022})}\BibitemShut {NoStop}%
\bibitem [{\citenamefont {Uvarov}\ \emph
  {et~al.}(2020{\natexlab{b}})\citenamefont {Uvarov}, \citenamefont
  {Biamonte},\ and\ \citenamefont {Yudin}}]{VQE_magnet}%
  \BibitemOpen
  \bibfield  {author} {\bibinfo {author} {\bibfnamefont {Alexey}\ \bibnamefont
  {Uvarov}}, \bibinfo {author} {\bibfnamefont {Jacob~D.}\ \bibnamefont
  {Biamonte}}, \ and\ \bibinfo {author} {\bibfnamefont {Dmitry}\ \bibnamefont
  {Yudin}},\ }\bibfield  {title} {\enquote {\bibinfo {title} {Variational
  quantum eigensolver for frustrated quantum systems},}\ }\href {\doibase
  10.1103/PhysRevB.102.075104} {\bibfield  {journal} {\bibinfo  {journal}
  {Phys. Rev. B}\ }\textbf {\bibinfo {volume} {102}},\ \bibinfo {pages}
  {075104} (\bibinfo {year} {2020}{\natexlab{b}})}\BibitemShut {NoStop}%
\bibitem [{\citenamefont {Dupont}\ and\ \citenamefont
  {Moore}(2022)}]{VQE_ising}%
  \BibitemOpen
  \bibfield  {author} {\bibinfo {author} {\bibfnamefont {Maxime}\ \bibnamefont
  {Dupont}}\ and\ \bibinfo {author} {\bibfnamefont {Joel~E.}\ \bibnamefont
  {Moore}},\ }\bibfield  {title} {\enquote {\bibinfo {title} {Quantum
  criticality using a superconducting quantum processor},}\ }\href {\doibase
  10.1103/PhysRevB.106.L041109} {\bibfield  {journal} {\bibinfo  {journal}
  {Phys. Rev. B}\ }\textbf {\bibinfo {volume} {106}},\ \bibinfo {pages}
  {L041109} (\bibinfo {year} {2022})}\BibitemShut {NoStop}%
\bibitem [{\citenamefont {Hlatshwayo}\ \emph {et~al.}(2022)\citenamefont
  {Hlatshwayo}, \citenamefont {Zhang}, \citenamefont {Wibowo}, \citenamefont
  {LaRose}, \citenamefont {Lacroix},\ and\ \citenamefont
  {Litvinova}}]{Hlatshwayo:2022yqt}%
  \BibitemOpen
  \bibfield  {author} {\bibinfo {author} {\bibfnamefont {Manqoba~Q.}\
  \bibnamefont {Hlatshwayo}}, \bibinfo {author} {\bibfnamefont {Yinu}\
  \bibnamefont {Zhang}}, \bibinfo {author} {\bibfnamefont {Herlik}\
  \bibnamefont {Wibowo}}, \bibinfo {author} {\bibfnamefont {Ryan}\ \bibnamefont
  {LaRose}}, \bibinfo {author} {\bibfnamefont {Denis}\ \bibnamefont {Lacroix}},
  \ and\ \bibinfo {author} {\bibfnamefont {Elena}\ \bibnamefont {Litvinova}},\
  }\bibfield  {title} {\enquote {\bibinfo {title} {{Simulating excited states
  of the {Lipkin} model on a quantum computer}},}\ }\href@noop {} {\  (\bibinfo
  {year} {2022})},\ \Eprint {http://arxiv.org/abs/2203.01478} {arXiv:2203.01478
  [nucl-th]} \BibitemShut {NoStop}%
\bibitem [{\citenamefont {Benedetti}\ \emph {et~al.}(2019)\citenamefont
  {Benedetti}, \citenamefont {Lloyd}, \citenamefont {Sack},\ and\ \citenamefont
  {Fiorentini}}]{PQC}%
  \BibitemOpen
  \bibfield  {author} {\bibinfo {author} {\bibfnamefont {Marcello}\
  \bibnamefont {Benedetti}}, \bibinfo {author} {\bibfnamefont {Erika}\
  \bibnamefont {Lloyd}}, \bibinfo {author} {\bibfnamefont {Stefan}\
  \bibnamefont {Sack}}, \ and\ \bibinfo {author} {\bibfnamefont {Mattia}\
  \bibnamefont {Fiorentini}},\ }\bibfield  {title} {\enquote {\bibinfo {title}
  {Parameterized quantum circuits as machine learning models},}\ }\href
  {\doibase 10.1088/2058-9565/ab4eb5} {\bibfield  {journal} {\bibinfo
  {journal} {Quantum Science and Technology}\ }\textbf {\bibinfo {volume}
  {4}},\ \bibinfo {pages} {4} (\bibinfo {year} {2019})}\BibitemShut {NoStop}%
\bibitem [{\citenamefont {Anand}\ \emph {et~al.}(2022)\citenamefont {Anand},
  \citenamefont {Schleich}, \citenamefont {Alperin-Lea}, \citenamefont
  {Jensen}, \citenamefont {Sim}, \citenamefont {D{\'\i}az-Tinoco},
  \citenamefont {Kottmann}, \citenamefont {Degroote}, \citenamefont
  {Izmaylov},\ and\ \citenamefont {Aspuru-Guzik}}]{anand2022quantum}%
  \BibitemOpen
  \bibfield  {author} {\bibinfo {author} {\bibfnamefont {Abhinav}\ \bibnamefont
  {Anand}}, \bibinfo {author} {\bibfnamefont {Philipp}\ \bibnamefont
  {Schleich}}, \bibinfo {author} {\bibfnamefont {Sumner}\ \bibnamefont
  {Alperin-Lea}}, \bibinfo {author} {\bibfnamefont {Phillip~WK}\ \bibnamefont
  {Jensen}}, \bibinfo {author} {\bibfnamefont {Sukin}\ \bibnamefont {Sim}},
  \bibinfo {author} {\bibfnamefont {Manuel}\ \bibnamefont {D{\'\i}az-Tinoco}},
  \bibinfo {author} {\bibfnamefont {Jakob~S}\ \bibnamefont {Kottmann}},
  \bibinfo {author} {\bibfnamefont {Matthias}\ \bibnamefont {Degroote}},
  \bibinfo {author} {\bibfnamefont {Artur~F}\ \bibnamefont {Izmaylov}}, \ and\
  \bibinfo {author} {\bibfnamefont {Al{\'a}n}\ \bibnamefont {Aspuru-Guzik}},\
  }\bibfield  {title} {\enquote {\bibinfo {title} {A quantum computing view on
  unitary coupled cluster theory},}\ }\href@noop {} {\bibfield  {journal}
  {\bibinfo  {journal} {Chemical Society Reviews}\ } (\bibinfo {year}
  {2022})}\BibitemShut {NoStop}%
\bibitem [{\citenamefont {Lee}\ \emph {et~al.}(2018)\citenamefont {Lee},
  \citenamefont {Huggins}, \citenamefont {Head-Gordon},\ and\ \citenamefont
  {Whaley}}]{lee2018generalized}%
  \BibitemOpen
  \bibfield  {author} {\bibinfo {author} {\bibfnamefont {Joonho}\ \bibnamefont
  {Lee}}, \bibinfo {author} {\bibfnamefont {William~J}\ \bibnamefont
  {Huggins}}, \bibinfo {author} {\bibfnamefont {Martin}\ \bibnamefont
  {Head-Gordon}}, \ and\ \bibinfo {author} {\bibfnamefont {K~Birgitta}\
  \bibnamefont {Whaley}},\ }\bibfield  {title} {\enquote {\bibinfo {title}
  {Generalized unitary coupled cluster wave functions for quantum
  computation},}\ }\href@noop {} {\bibfield  {journal} {\bibinfo  {journal}
  {Journal of chemical theory and computation}\ }\textbf {\bibinfo {volume}
  {15}},\ \bibinfo {pages} {311--324} (\bibinfo {year} {2018})}\BibitemShut
  {NoStop}%
\bibitem [{\citenamefont {McClean}\ \emph {et~al.}(2018)\citenamefont
  {McClean}, \citenamefont {Boixo}, \citenamefont {Smelyanskiy}, \citenamefont
  {Babbush},\ and\ \citenamefont {Neven}}]{Barren_platea_McClean}%
  \BibitemOpen
  \bibfield  {author} {\bibinfo {author} {\bibfnamefont {Jarrod~R.}\
  \bibnamefont {McClean}}, \bibinfo {author} {\bibfnamefont {Sergio}\
  \bibnamefont {Boixo}}, \bibinfo {author} {\bibfnamefont {Vadim~N.}\
  \bibnamefont {Smelyanskiy}}, \bibinfo {author} {\bibfnamefont {Ryan}\
  \bibnamefont {Babbush}}, \ and\ \bibinfo {author} {\bibfnamefont {Hartmut}\
  \bibnamefont {Neven}},\ }\bibfield  {title} {\enquote {\bibinfo {title}
  {Barren plateaus in quantum neural network training landscapes},}\ }\href
  {\doibase https://doi.org/10.1038/s41467-018-07090-4} {\bibfield  {journal}
  {\bibinfo  {journal} {Nature Communications}\ }\textbf {\bibinfo {volume}
  {9}} (\bibinfo {year} {2018}),\
  https://doi.org/10.1038/s41467-018-07090-4}\BibitemShut {NoStop}%
\bibitem [{\citenamefont {Schatzki}\ \emph {et~al.}(2022)\citenamefont
  {Schatzki}, \citenamefont {Larocca}, \citenamefont {Nguyen}, \citenamefont
  {Sauvage},\ and\ \citenamefont {Cerezo}}]{equivar}%
  \BibitemOpen
  \bibfield  {author} {\bibinfo {author} {\bibfnamefont {Louis}\ \bibnamefont
  {Schatzki}}, \bibinfo {author} {\bibfnamefont {Martin}\ \bibnamefont
  {Larocca}}, \bibinfo {author} {\bibfnamefont {Quynh~T.}\ \bibnamefont
  {Nguyen}}, \bibinfo {author} {\bibfnamefont {Frederic}\ \bibnamefont
  {Sauvage}}, \ and\ \bibinfo {author} {\bibfnamefont {M.}~\bibnamefont
  {Cerezo}},\ }\href {\doibase 10.48550/ARXIV.2210.09974} {\enquote {\bibinfo
  {title} {Theoretical guarantees for permutation-equivariant quantum neural
  networks},}\ } (\bibinfo {year} {2022})\BibitemShut {NoStop}%
\bibitem [{\citenamefont {Grimsley}\ \emph {et~al.}(2019)\citenamefont
  {Grimsley}, \citenamefont {Economou}, \citenamefont {Barnes},\ and\
  \citenamefont {Mayhall}}]{ADAPT-VQE}%
  \BibitemOpen
  \bibfield  {author} {\bibinfo {author} {\bibfnamefont {Harper~R.}\
  \bibnamefont {Grimsley}}, \bibinfo {author} {\bibfnamefont {Sophia~E.}\
  \bibnamefont {Economou}}, \bibinfo {author} {\bibfnamefont {Edwin}\
  \bibnamefont {Barnes}}, \ and\ \bibinfo {author} {\bibfnamefont
  {Nicholas~J.}\ \bibnamefont {Mayhall}},\ }\bibfield  {title} {\enquote
  {\bibinfo {title} {An adaptive variational algorithm for exact molecular
  simulations on a quantum computer},}\ }\href {\doibase
  https://doi.org/10.1038/s41467-019-10988-2} {\bibfield  {journal} {\bibinfo
  {journal} {Nat Commun}\ }\textbf {\bibinfo {volume} {10}},\ \bibinfo {pages}
  {3007} (\bibinfo {year} {2019})}\BibitemShut {NoStop}%
\bibitem [{\citenamefont {Higgott}\ \emph {et~al.}(2019)\citenamefont
  {Higgott}, \citenamefont {Wang},\ and\ \citenamefont
  {Brierley}}]{Excited-states}%
  \BibitemOpen
  \bibfield  {author} {\bibinfo {author} {\bibfnamefont {Oscar}\ \bibnamefont
  {Higgott}}, \bibinfo {author} {\bibfnamefont {Daochen}\ \bibnamefont {Wang}},
  \ and\ \bibinfo {author} {\bibfnamefont {Stephen}\ \bibnamefont {Brierley}},\
  }\bibfield  {title} {\enquote {\bibinfo {title} {Variational quantum
  computation of excited states},}\ }\href {\doibase
  https://doi.org/10.22331/q-2019-07-01-156} {\bibfield  {journal} {\bibinfo
  {journal} {Quantum}\ }\textbf {\bibinfo {volume} {3}},\ \bibinfo {pages}
  {156} (\bibinfo {year} {2019})}\BibitemShut {NoStop}%
\bibitem [{\citenamefont {Ollitrault}\ \emph {et~al.}(2020)\citenamefont
  {Ollitrault}, \citenamefont {Kandala}, \citenamefont {Chen}, \citenamefont
  {Barkoutsos}, \citenamefont {Mezzacapo}, \citenamefont {Pistoia},
  \citenamefont {Sheldon}, \citenamefont {Woerner}, \citenamefont {Gambetta},\
  and\ \citenamefont {Tavernelli}}]{Ollitrault_excited_state}%
  \BibitemOpen
  \bibfield  {author} {\bibinfo {author} {\bibfnamefont {Pauline~J.}\
  \bibnamefont {Ollitrault}}, \bibinfo {author} {\bibfnamefont {Abhinav}\
  \bibnamefont {Kandala}}, \bibinfo {author} {\bibfnamefont {Chun-Fu}\
  \bibnamefont {Chen}}, \bibinfo {author} {\bibfnamefont {Panagiotis~Kl.}\
  \bibnamefont {Barkoutsos}}, \bibinfo {author} {\bibfnamefont {Antonio}\
  \bibnamefont {Mezzacapo}}, \bibinfo {author} {\bibfnamefont {Marco}\
  \bibnamefont {Pistoia}}, \bibinfo {author} {\bibfnamefont {Sarah}\
  \bibnamefont {Sheldon}}, \bibinfo {author} {\bibfnamefont {Stefan}\
  \bibnamefont {Woerner}}, \bibinfo {author} {\bibfnamefont {Jay~M.}\
  \bibnamefont {Gambetta}}, \ and\ \bibinfo {author} {\bibfnamefont {Ivano}\
  \bibnamefont {Tavernelli}},\ }\bibfield  {title} {\enquote {\bibinfo {title}
  {Quantum equation of motion for computing molecular excitation energies on a
  noisy quantum processor},}\ }\href {\doibase
  10.1103/PhysRevResearch.2.043140} {\bibfield  {journal} {\bibinfo  {journal}
  {Phys. Rev. Research}\ }\textbf {\bibinfo {volume} {2}},\ \bibinfo {pages}
  {043140} (\bibinfo {year} {2020})}\BibitemShut {NoStop}%
\bibitem [{\citenamefont {Tilly}\ \emph {et~al.}(2020)\citenamefont {Tilly},
  \citenamefont {Jones}, \citenamefont {Chen}, \citenamefont {Wossnig},\ and\
  \citenamefont {Grant}}]{DVQE}%
  \BibitemOpen
  \bibfield  {author} {\bibinfo {author} {\bibfnamefont {Jules}\ \bibnamefont
  {Tilly}}, \bibinfo {author} {\bibfnamefont {Glenn}\ \bibnamefont {Jones}},
  \bibinfo {author} {\bibfnamefont {Hongxiang}\ \bibnamefont {Chen}}, \bibinfo
  {author} {\bibfnamefont {Leonard}\ \bibnamefont {Wossnig}}, \ and\ \bibinfo
  {author} {\bibfnamefont {Edward}\ \bibnamefont {Grant}},\ }\bibfield  {title}
  {\enquote {\bibinfo {title} {Computation of molecular excited states on ibm
  quantum computers using a discriminative variational quantum eigensolver},}\
  }\href {\doibase 10.1103/PhysRevA.102.062425} {\bibfield  {journal} {\bibinfo
   {journal} {Phys. Rev. A}\ }\textbf {\bibinfo {volume} {102}},\ \bibinfo
  {pages} {062425} (\bibinfo {year} {2020})}\BibitemShut {NoStop}%
\bibitem [{\citenamefont {Nakanishi}\ \emph {et~al.}(2019)\citenamefont
  {Nakanishi}, \citenamefont {Mitarai},\ and\ \citenamefont
  {Fujii}}]{SSVQE_Mitarai}%
  \BibitemOpen
  \bibfield  {author} {\bibinfo {author} {\bibfnamefont {Ken~M.}\ \bibnamefont
  {Nakanishi}}, \bibinfo {author} {\bibfnamefont {Kosuke}\ \bibnamefont
  {Mitarai}}, \ and\ \bibinfo {author} {\bibfnamefont {Keisuke}\ \bibnamefont
  {Fujii}},\ }\bibfield  {title} {\enquote {\bibinfo {title} {Subspace-search
  variational quantum eigensolver for excited states},}\ }\href {\doibase
  10.1103/PhysRevResearch.1.033062} {\bibfield  {journal} {\bibinfo  {journal}
  {Phys. Rev. Research}\ }\textbf {\bibinfo {volume} {1}},\ \bibinfo {pages}
  {033062} (\bibinfo {year} {2019})}\BibitemShut {NoStop}%
\bibitem [{\citenamefont {Nation}\ \emph {et~al.}(2021)\citenamefont {Nation},
  \citenamefont {Kang}, \citenamefont {Sundaresan},\ and\ \citenamefont
  {Gambetta}}]{MEM}%
  \BibitemOpen
  \bibfield  {author} {\bibinfo {author} {\bibfnamefont {Paul~D.}\ \bibnamefont
  {Nation}}, \bibinfo {author} {\bibfnamefont {Hwajung}\ \bibnamefont {Kang}},
  \bibinfo {author} {\bibfnamefont {Neereja}\ \bibnamefont {Sundaresan}}, \
  and\ \bibinfo {author} {\bibfnamefont {Jay~M.}\ \bibnamefont {Gambetta}},\
  }\bibfield  {title} {\enquote {\bibinfo {title} {Scalable mitigation of
  measurement errors on quantum computers},}\ }\href {\doibase
  10.1103/PRXQuantum.2.040326} {\bibfield  {journal} {\bibinfo  {journal} {PRX
  Quantum}\ }\textbf {\bibinfo {volume} {2}},\ \bibinfo {pages} {040326}
  (\bibinfo {year} {2021})}\BibitemShut {NoStop}%
\bibitem [{\citenamefont {He}\ \emph {et~al.}(2020)\citenamefont {He},
  \citenamefont {Nachman}, \citenamefont {de~Jong},\ and\ \citenamefont
  {Bauer}}]{ZNE}%
  \BibitemOpen
  \bibfield  {author} {\bibinfo {author} {\bibfnamefont {Andre}\ \bibnamefont
  {He}}, \bibinfo {author} {\bibfnamefont {Benjamin}\ \bibnamefont {Nachman}},
  \bibinfo {author} {\bibfnamefont {Wibe~A.}\ \bibnamefont {de~Jong}}, \ and\
  \bibinfo {author} {\bibfnamefont {Christian~W.}\ \bibnamefont {Bauer}},\
  }\bibfield  {title} {\enquote {\bibinfo {title} {Zero-noise extrapolation for
  quantum-gate error mitigation with identity insertions},}\ }\href {\doibase
  10.1103/PhysRevA.102.012426} {\bibfield  {journal} {\bibinfo  {journal}
  {Phys. Rev. A}\ }\textbf {\bibinfo {volume} {102}},\ \bibinfo {pages}
  {012426} (\bibinfo {year} {2020})}\BibitemShut {NoStop}%
\bibitem [{\citenamefont {Kandala}\ \emph {et~al.}(2019)\citenamefont
  {Kandala}, \citenamefont {Temme}, \citenamefont {C\'orcoles}, \citenamefont
  {Mezzacapo}, \citenamefont {Chow},\ and\ \citenamefont
  {Gambetta}}]{error-mitigation}%
  \BibitemOpen
  \bibfield  {author} {\bibinfo {author} {\bibfnamefont {Abhinav}\ \bibnamefont
  {Kandala}}, \bibinfo {author} {\bibfnamefont {Kristan}\ \bibnamefont
  {Temme}}, \bibinfo {author} {\bibfnamefont {Antonio~D.}\ \bibnamefont
  {C\'orcoles}}, \bibinfo {author} {\bibfnamefont {Antonio}\ \bibnamefont
  {Mezzacapo}}, \bibinfo {author} {\bibfnamefont {Jerry~M.}\ \bibnamefont
  {Chow}}, \ and\ \bibinfo {author} {\bibfnamefont {Jay~M.}\ \bibnamefont
  {Gambetta}},\ }\bibfield  {title} {\enquote {\bibinfo {title} {Error
  mitigation extends the computational reach of a noisy quantum processor},}\
  }\href {\doibase https://doi.org/10.1038/s41586-019-1040-7} {\bibfield
  {journal} {\bibinfo  {journal} {Nature}\ }\textbf {\bibinfo {volume} {567}},\
  \bibinfo {pages} {491–495} (\bibinfo {year} {2019})}\BibitemShut {NoStop}%
\bibitem [{\citenamefont {Richardson}(1911)}]{Richardson}%
  \BibitemOpen
  \bibfield  {author} {\bibinfo {author} {\bibfnamefont {Lewis~Fry}\
  \bibnamefont {Richardson}},\ }\bibfield  {title} {\enquote {\bibinfo {title}
  {The approximate arithmetical solution by finite differences of physical
  problems involving differential equations, with an application to the
  stresses in a masonry dam},}\ }\href {\doibase
  http://doi.org/10.1098/rsta.1911.0009} {\bibfield  {journal} {\bibinfo
  {journal} {Philosophical Transactions of the Royal Society of London, Series
  A, Containing Papers of a Mathematical or Physical Character}\ }\textbf
  {\bibinfo {volume} {202}},\ \bibinfo {pages} {307–357} (\bibinfo {year}
  {1911})}\BibitemShut {NoStop}%
\bibitem [{\citenamefont {Takagi}\ \emph
  {et~al.}(2022{\natexlab{a}})\citenamefont {Takagi}, \citenamefont {Endo},
  \citenamefont {Minagawa},\ and\ \citenamefont {Gu}}]{Takagi_2022}%
  \BibitemOpen
  \bibfield  {author} {\bibinfo {author} {\bibfnamefont {Ryuji}\ \bibnamefont
  {Takagi}}, \bibinfo {author} {\bibfnamefont {Suguru}\ \bibnamefont {Endo}},
  \bibinfo {author} {\bibfnamefont {Shintaro}\ \bibnamefont {Minagawa}}, \ and\
  \bibinfo {author} {\bibfnamefont {Mile}\ \bibnamefont {Gu}},\ }\bibfield
  {title} {\enquote {\bibinfo {title} {Fundamental limits of quantum error
  mitigation},}\ }\href {\doibase 10.1038/s41534-022-00618-z} {\bibfield
  {journal} {\bibinfo  {journal} {npj Quantum Information}\ }\textbf {\bibinfo
  {volume} {8}} (\bibinfo {year} {2022}{\natexlab{a}}),\
  10.1038/s41534-022-00618-z}\BibitemShut {NoStop}%
\bibitem [{\citenamefont {Lieb}\ \emph {et~al.}(1961)\citenamefont {Lieb},
  \citenamefont {Schultz},\ and\ \citenamefont {Mattis}}]{lieb1961}%
  \BibitemOpen
  \bibfield  {author} {\bibinfo {author} {\bibfnamefont {Elliott}\ \bibnamefont
  {Lieb}}, \bibinfo {author} {\bibfnamefont {Theodore}\ \bibnamefont
  {Schultz}}, \ and\ \bibinfo {author} {\bibfnamefont {Daniel}\ \bibnamefont
  {Mattis}},\ }\bibfield  {title} {\enquote {\bibinfo {title} {Two soluble
  models of an antiferromagnetic chain},}\ }\href@noop {} {\bibfield  {journal}
  {\bibinfo  {journal} {Annals of Physics}\ }\textbf {\bibinfo {volume} {16}},\
  \bibinfo {pages} {407--466} (\bibinfo {year} {1961})}\BibitemShut {NoStop}%
\bibitem [{\citenamefont {Dusuel}\ and\ \citenamefont {Vidal}(2005)}]{LMG_MB1}%
  \BibitemOpen
  \bibfield  {author} {\bibinfo {author} {\bibfnamefont {S\'ebastien}\
  \bibnamefont {Dusuel}}\ and\ \bibinfo {author} {\bibfnamefont {Julien}\
  \bibnamefont {Vidal}},\ }\bibfield  {title} {\enquote {\bibinfo {title}
  {Continuous unitary transformations and finite-size scaling exponents in the
  {Lipkin}-{Meshkov}-{Glick} model},}\ }\href {\doibase
  10.1103/PhysRevB.71.224420} {\bibfield  {journal} {\bibinfo  {journal} {Phys.
  Rev. B}\ }\textbf {\bibinfo {volume} {71}},\ \bibinfo {pages} {224420}
  (\bibinfo {year} {2005})}\BibitemShut {NoStop}%
\bibitem [{\citenamefont {Cirac}\ \emph {et~al.}(1998)\citenamefont {Cirac},
  \citenamefont {Lewenstein}, \citenamefont {M\o{}lmer},\ and\ \citenamefont
  {Zoller}}]{LMG_impl1}%
  \BibitemOpen
  \bibfield  {author} {\bibinfo {author} {\bibfnamefont {J.~I.}\ \bibnamefont
  {Cirac}}, \bibinfo {author} {\bibfnamefont {M.}~\bibnamefont {Lewenstein}},
  \bibinfo {author} {\bibfnamefont {K.}~\bibnamefont {M\o{}lmer}}, \ and\
  \bibinfo {author} {\bibfnamefont {P.}~\bibnamefont {Zoller}},\ }\bibfield
  {title} {\enquote {\bibinfo {title} {Quantum superposition states of
  {Bose}-{Einstein} condensates},}\ }\href {\doibase 10.1103/PhysRevA.57.1208}
  {\bibfield  {journal} {\bibinfo  {journal} {Phys. Rev. A}\ }\textbf {\bibinfo
  {volume} {57}},\ \bibinfo {pages} {1208--1218} (\bibinfo {year}
  {1998})}\BibitemShut {NoStop}%
\bibitem [{\citenamefont {Garanin}\ \emph {et~al.}(1998)\citenamefont
  {Garanin}, \citenamefont {Mart\'{\i}nez~Hidalgo},\ and\ \citenamefont
  {Chudnovsky}}]{LMG_impl2}%
  \BibitemOpen
  \bibfield  {author} {\bibinfo {author} {\bibfnamefont {D.~A.}\ \bibnamefont
  {Garanin}}, \bibinfo {author} {\bibfnamefont {X.}~\bibnamefont
  {Mart\'{\i}nez~Hidalgo}}, \ and\ \bibinfo {author} {\bibfnamefont {E.~M.}\
  \bibnamefont {Chudnovsky}},\ }\bibfield  {title} {\enquote {\bibinfo {title}
  {Quantum-classical transition of the escape rate of a uniaxial spin system in
  an arbitrarily directed field},}\ }\href {\doibase 10.1103/PhysRevB.57.13639}
  {\bibfield  {journal} {\bibinfo  {journal} {Phys. Rev. B}\ }\textbf {\bibinfo
  {volume} {57}},\ \bibinfo {pages} {13639--13654} (\bibinfo {year}
  {1998})}\BibitemShut {NoStop}%
\bibitem [{\citenamefont {Casta\~nos}\ \emph {et~al.}(2006)\citenamefont
  {Casta\~nos}, \citenamefont {L\'opez-Pe\~na}, \citenamefont {Hirsch},\ and\
  \citenamefont {L\'opez-Moreno}}]{LMG_impl3}%
  \BibitemOpen
  \bibfield  {author} {\bibinfo {author} {\bibfnamefont {Octavio}\ \bibnamefont
  {Casta\~nos}}, \bibinfo {author} {\bibfnamefont {Ram\'on}\ \bibnamefont
  {L\'opez-Pe\~na}}, \bibinfo {author} {\bibfnamefont {Jorge~G.}\ \bibnamefont
  {Hirsch}}, \ and\ \bibinfo {author} {\bibfnamefont {Enrique}\ \bibnamefont
  {L\'opez-Moreno}},\ }\bibfield  {title} {\enquote {\bibinfo {title}
  {Classical and quantum phase transitions in the {Lipkin}-{Meshkov}-{Glick}
  model},}\ }\href {\doibase 10.1103/PhysRevB.74.104118} {\bibfield  {journal}
  {\bibinfo  {journal} {Phys. Rev. B}\ }\textbf {\bibinfo {volume} {74}},\
  \bibinfo {pages} {104118} (\bibinfo {year} {2006})}\BibitemShut {NoStop}%
\bibitem [{\citenamefont {Ma}\ and\ \citenamefont {Wang}(2009)}]{LMG_impl4}%
  \BibitemOpen
  \bibfield  {author} {\bibinfo {author} {\bibfnamefont {Jian}\ \bibnamefont
  {Ma}}\ and\ \bibinfo {author} {\bibfnamefont {Xiaoguang}\ \bibnamefont
  {Wang}},\ }\bibfield  {title} {\enquote {\bibinfo {title} {Fisher information
  and spin squeezing in the {Lipkin}-{Meshkov}-{Glick} model},}\ }\href
  {\doibase 10.1103/PhysRevA.80.012318} {\bibfield  {journal} {\bibinfo
  {journal} {Phys. Rev. A}\ }\textbf {\bibinfo {volume} {80}},\ \bibinfo
  {pages} {012318} (\bibinfo {year} {2009})}\BibitemShut {NoStop}%
\bibitem [{\citenamefont {Vidal}\ \emph {et~al.}(2004)\citenamefont {Vidal},
  \citenamefont {Palacios},\ and\ \citenamefont {Aslangul}}]{LMG_QT1}%
  \BibitemOpen
  \bibfield  {author} {\bibinfo {author} {\bibfnamefont {Julien}\ \bibnamefont
  {Vidal}}, \bibinfo {author} {\bibfnamefont {Guillaume}\ \bibnamefont
  {Palacios}}, \ and\ \bibinfo {author} {\bibfnamefont {Claude}\ \bibnamefont
  {Aslangul}},\ }\bibfield  {title} {\enquote {\bibinfo {title} {Entanglement
  dynamics in the {Lipkin}-{Meshkov}-{Glick} model},}\ }\href {\doibase
  10.1103/PhysRevA.70.062304} {\bibfield  {journal} {\bibinfo  {journal} {Phys.
  Rev. A}\ }\textbf {\bibinfo {volume} {70}},\ \bibinfo {pages} {062304}
  (\bibinfo {year} {2004})}\BibitemShut {NoStop}%
\bibitem [{\citenamefont {Dusuel}\ and\ \citenamefont {Vidal}(2004)}]{LMG_QT2}%
  \BibitemOpen
  \bibfield  {author} {\bibinfo {author} {\bibfnamefont {S\'ebastien}\
  \bibnamefont {Dusuel}}\ and\ \bibinfo {author} {\bibfnamefont {Julien}\
  \bibnamefont {Vidal}},\ }\bibfield  {title} {\enquote {\bibinfo {title}
  {Finite-size scaling exponents of the {Lipkin}-{Meshkov}-{Glick} model},}\
  }\href {\doibase 10.1103/PhysRevLett.93.237204} {\bibfield  {journal}
  {\bibinfo  {journal} {Phys. Rev. Lett.}\ }\textbf {\bibinfo {volume} {93}},\
  \bibinfo {pages} {237204} (\bibinfo {year} {2004})}\BibitemShut {NoStop}%
\bibitem [{\citenamefont {Ribeiro}\ \emph {et~al.}(2008)\citenamefont
  {Ribeiro}, \citenamefont {Vidal},\ and\ \citenamefont {Mosseri}}]{LMG_QT3}%
  \BibitemOpen
  \bibfield  {author} {\bibinfo {author} {\bibfnamefont {Pedro}\ \bibnamefont
  {Ribeiro}}, \bibinfo {author} {\bibfnamefont {Julien}\ \bibnamefont {Vidal}},
  \ and\ \bibinfo {author} {\bibfnamefont {R\'emy}\ \bibnamefont {Mosseri}},\
  }\bibfield  {title} {\enquote {\bibinfo {title} {Exact spectrum of the
  {Lipkin}-{Meshkov}-{Glick} model in the thermodynamic limit and finite-size
  corrections},}\ }\href {\doibase 10.1103/PhysRevE.78.021106} {\bibfield
  {journal} {\bibinfo  {journal} {Phys. Rev. E}\ }\textbf {\bibinfo {volume}
  {78}},\ \bibinfo {pages} {021106} (\bibinfo {year} {2008})}\BibitemShut
  {NoStop}%
\bibitem [{\citenamefont {Salvatori}\ \emph {et~al.}(2014)\citenamefont
  {Salvatori}, \citenamefont {Mandarino},\ and\ \citenamefont
  {Paris}}]{LMG_QT6}%
  \BibitemOpen
  \bibfield  {author} {\bibinfo {author} {\bibfnamefont {Giulio}\ \bibnamefont
  {Salvatori}}, \bibinfo {author} {\bibfnamefont {Antonio}\ \bibnamefont
  {Mandarino}}, \ and\ \bibinfo {author} {\bibfnamefont {Matteo G.~A.}\
  \bibnamefont {Paris}},\ }\bibfield  {title} {\enquote {\bibinfo {title}
  {Quantum metrology in {Lipkin}-{Meshkov}-{Glick} critical systems},}\ }\href
  {\doibase 10.1103/PhysRevA.90.022111} {\bibfield  {journal} {\bibinfo
  {journal} {Phys. Rev. A}\ }\textbf {\bibinfo {volume} {90}},\ \bibinfo
  {pages} {022111} (\bibinfo {year} {2014})}\BibitemShut {NoStop}%
\bibitem [{\citenamefont {Mandarino}\ \emph {et~al.}(2021)\citenamefont
  {Mandarino}, \citenamefont {Joulain}, \citenamefont {G\'omez},\ and\
  \citenamefont {Bellomo}}]{LMG_QT7}%
  \BibitemOpen
  \bibfield  {author} {\bibinfo {author} {\bibfnamefont {Antonio}\ \bibnamefont
  {Mandarino}}, \bibinfo {author} {\bibfnamefont {Karl}\ \bibnamefont
  {Joulain}}, \bibinfo {author} {\bibfnamefont {Melisa~Dom\'{\i}nguez}\
  \bibnamefont {G\'omez}}, \ and\ \bibinfo {author} {\bibfnamefont {Bruno}\
  \bibnamefont {Bellomo}},\ }\bibfield  {title} {\enquote {\bibinfo {title}
  {Thermal transistor effect in quantum systems},}\ }\href {\doibase
  10.1103/PhysRevApplied.16.034026} {\bibfield  {journal} {\bibinfo  {journal}
  {Phys. Rev. Applied}\ }\textbf {\bibinfo {volume} {16}},\ \bibinfo {pages}
  {034026} (\bibinfo {year} {2021})}\BibitemShut {NoStop}%
\bibitem [{\citenamefont {Mandarino}(2022)}]{LMG_QT8}%
  \BibitemOpen
  \bibfield  {author} {\bibinfo {author} {\bibfnamefont {Antonio}\ \bibnamefont
  {Mandarino}},\ }\bibfield  {title} {\enquote {\bibinfo {title} {Quantum
  thermal amplifiers with engineered dissipation},}\ }\href {\doibase
  10.3390/e24081031} {\bibfield  {journal} {\bibinfo  {journal} {Entropy}\
  }\textbf {\bibinfo {volume} {24}} (\bibinfo {year} {2022}),\
  10.3390/e24081031}\BibitemShut {NoStop}%
\bibitem [{\citenamefont {Agassi}\ \emph {et~al.}(1966)\citenamefont {Agassi},
  \citenamefont {{Lipkin}},\ and\ \citenamefont {{Meshkov}}}]{LMG4}%
  \BibitemOpen
  \bibfield  {author} {\bibinfo {author} {\bibfnamefont {D}~\bibnamefont
  {Agassi}}, \bibinfo {author} {\bibfnamefont {HJ}~\bibnamefont {{Lipkin}}}, \
  and\ \bibinfo {author} {\bibfnamefont {N}~\bibnamefont {{Meshkov}}},\
  }\bibfield  {title} {\enquote {\bibinfo {title} {Validity of many-body
  approximation methods for a solvable model:(iv). the deformed hartree-fock
  solution},}\ }\href@noop {} {\bibfield  {journal} {\bibinfo  {journal}
  {Nuclear Physics}\ }\textbf {\bibinfo {volume} {86}},\ \bibinfo {pages}
  {321--331} (\bibinfo {year} {1966})}\BibitemShut {NoStop}%
\bibitem [{\citenamefont {Dicke}(1954)}]{Dicke54}%
  \BibitemOpen
  \bibfield  {author} {\bibinfo {author} {\bibfnamefont {R.~H.}\ \bibnamefont
  {Dicke}},\ }\bibfield  {title} {\enquote {\bibinfo {title} {Coherence in
  spontaneous radiation processes},}\ }\href {\doibase 10.1103/PhysRev.93.99}
  {\bibfield  {journal} {\bibinfo  {journal} {Phys. Rev.}\ }\textbf {\bibinfo
  {volume} {93}},\ \bibinfo {pages} {99--110} (\bibinfo {year}
  {1954})}\BibitemShut {NoStop}%
\bibitem [{\citenamefont {Das}\ \emph {et~al.}(2006)\citenamefont {Das},
  \citenamefont {Sengupta}, \citenamefont {Sen},\ and\ \citenamefont
  {Chakrabarti}}]{Ising_infinite}%
  \BibitemOpen
  \bibfield  {author} {\bibinfo {author} {\bibfnamefont {Arnab}\ \bibnamefont
  {Das}}, \bibinfo {author} {\bibfnamefont {K.}~\bibnamefont {Sengupta}},
  \bibinfo {author} {\bibfnamefont {Diptiman}\ \bibnamefont {Sen}}, \ and\
  \bibinfo {author} {\bibfnamefont {Bikas~K.}\ \bibnamefont {Chakrabarti}},\
  }\bibfield  {title} {\enquote {\bibinfo {title} {Infinite-range {Ising}
  ferromagnet in a time-dependent transverse magnetic field: Quench and ac
  dynamics near the quantum critical point},}\ }\href {\doibase
  10.1103/PhysRevB.74.144423} {\bibfield  {journal} {\bibinfo  {journal} {Phys.
  Rev. B}\ }\textbf {\bibinfo {volume} {74}},\ \bibinfo {pages} {144423}
  (\bibinfo {year} {2006})}\BibitemShut {NoStop}%
\bibitem [{\citenamefont {{Lerma H.}}\ and\ \citenamefont
  {Dukelsky}(2013)}]{LMG_gaudin}%
  \BibitemOpen
  \bibfield  {author} {\bibinfo {author} {\bibfnamefont {S.}~\bibnamefont
  {{Lerma H.}}}\ and\ \bibinfo {author} {\bibfnamefont {J.}~\bibnamefont
  {Dukelsky}},\ }\bibfield  {title} {\enquote {\bibinfo {title} {The
  {Lipkin}–{Meshkov}–{Glick} model as a particular limit of the su(1,1)
  {Richardson}–{Gaudin} integrable models},}\ }\href {\doibase
  https://doi.org/10.1016/j.nuclphysb.2013.01.019} {\bibfield  {journal}
  {\bibinfo  {journal} {Nuclear Physics B}\ }\textbf {\bibinfo {volume}
  {870}},\ \bibinfo {pages} {421--443} (\bibinfo {year} {2013})}\BibitemShut
  {NoStop}%
\bibitem [{\citenamefont {Chen}\ and\ \citenamefont {Liang}(2006)}]{LMG_MB2}%
  \BibitemOpen
  \bibfield  {author} {\bibinfo {author} {\bibfnamefont {Gang}\ \bibnamefont
  {Chen}}\ and\ \bibinfo {author} {\bibfnamefont {JQ}~\bibnamefont {Liang}},\
  }\bibfield  {title} {\enquote {\bibinfo {title} {Unconventional quantum phase
  transition in the finite-size {Lipkin}--{Meshkov}--{Glick} model},}\
  }\href@noop {} {\bibfield  {journal} {\bibinfo  {journal} {New Journal of
  Physics}\ }\textbf {\bibinfo {volume} {8}},\ \bibinfo {pages} {297} (\bibinfo
  {year} {2006})}\BibitemShut {NoStop}%
\bibitem [{\citenamefont {Kraft}(1988{\natexlab{a}})}]{kraft1988software}%
  \BibitemOpen
  \bibfield  {author} {\bibinfo {author} {\bibfnamefont {D.}~\bibnamefont
  {Kraft}},\ }\href {https://books.google.co.uk/books?id=4rKaGwAACAAJ} {\emph
  {\bibinfo {title} {A Software Package for Sequential Quadratic
  Programming}}},\ Deutsche Forschungs- und Versuchsanstalt f{\"u}r Luft- und
  Raumfahrt K{\"o}ln: Forschungsbericht\ (\bibinfo  {publisher} {Wiss.
  Berichtswesen d. DFVLR},\ \bibinfo {year} {1988})\BibitemShut {NoStop}%
\bibitem [{\citenamefont {Harwood}\ \emph {et~al.}(2022)\citenamefont
  {Harwood}, \citenamefont {Trenev}, \citenamefont {Stober}, \citenamefont
  {Barkoutsos}, \citenamefont {Gujarati}, \citenamefont {Mostame},\ and\
  \citenamefont {Greenberg}}]{adiabatic_VQE}%
  \BibitemOpen
  \bibfield  {author} {\bibinfo {author} {\bibfnamefont {Stuart~M.}\
  \bibnamefont {Harwood}}, \bibinfo {author} {\bibfnamefont {Dimitar}\
  \bibnamefont {Trenev}}, \bibinfo {author} {\bibfnamefont {Spencer~T.}\
  \bibnamefont {Stober}}, \bibinfo {author} {\bibfnamefont {Panagiotis}\
  \bibnamefont {Barkoutsos}}, \bibinfo {author} {\bibfnamefont {Tanvi~P.}\
  \bibnamefont {Gujarati}}, \bibinfo {author} {\bibfnamefont {Sarah}\
  \bibnamefont {Mostame}}, \ and\ \bibinfo {author} {\bibfnamefont {Donny}\
  \bibnamefont {Greenberg}},\ }\bibfield  {title} {\enquote {\bibinfo {title}
  {Improving the variational quantum eigensolver using variational adiabatic
  quantum computing},}\ }\href {\doibase 10.1145/3479197} {\bibfield  {journal}
  {\bibinfo  {journal} {ACM Transactions on Quantum Computing}\ }\textbf
  {\bibinfo {volume} {3}} (\bibinfo {year} {2022}),\
  10.1145/3479197}\BibitemShut {NoStop}%
\bibitem [{\citenamefont {Chow}\ \emph {et~al.}(2011)\citenamefont {Chow},
  \citenamefont {C\'orcoles}, \citenamefont {Gambetta}, \citenamefont
  {Rigetti}, \citenamefont {Johnson}, \citenamefont {Smolin}, \citenamefont
  {Rozen}, \citenamefont {Keefe}, \citenamefont {Rothwell}, \citenamefont
  {Ketchen},\ and\ \citenamefont {Steffen}}]{cross-resonance}%
  \BibitemOpen
  \bibfield  {author} {\bibinfo {author} {\bibfnamefont {Jerry~M.}\
  \bibnamefont {Chow}}, \bibinfo {author} {\bibfnamefont {A.~D.}\ \bibnamefont
  {C\'orcoles}}, \bibinfo {author} {\bibfnamefont {Jay~M.}\ \bibnamefont
  {Gambetta}}, \bibinfo {author} {\bibfnamefont {Chad}\ \bibnamefont
  {Rigetti}}, \bibinfo {author} {\bibfnamefont {B.~R.}\ \bibnamefont
  {Johnson}}, \bibinfo {author} {\bibfnamefont {John~A.}\ \bibnamefont
  {Smolin}}, \bibinfo {author} {\bibfnamefont {J.~R.}\ \bibnamefont {Rozen}},
  \bibinfo {author} {\bibfnamefont {George~A.}\ \bibnamefont {Keefe}}, \bibinfo
  {author} {\bibfnamefont {Mary~B.}\ \bibnamefont {Rothwell}}, \bibinfo
  {author} {\bibfnamefont {Mark~B.}\ \bibnamefont {Ketchen}}, \ and\ \bibinfo
  {author} {\bibfnamefont {M.}~\bibnamefont {Steffen}},\ }\bibfield  {title}
  {\enquote {\bibinfo {title} {Simple all-microwave entangling gate for
  fixed-frequency superconducting qubits},}\ }\href {\doibase
  10.1103/PhysRevLett.107.080502} {\bibfield  {journal} {\bibinfo  {journal}
  {Phys. Rev. Lett.}\ }\textbf {\bibinfo {volume} {107}},\ \bibinfo {pages}
  {080502} (\bibinfo {year} {2011})}\BibitemShut {NoStop}%
\bibitem [{\citenamefont {Li}\ \emph {et~al.}(2019)\citenamefont {Li},
  \citenamefont {Ding},\ and\ \citenamefont {Xie}}]{sabre}%
  \BibitemOpen
  \bibfield  {author} {\bibinfo {author} {\bibfnamefont {Gushu}\ \bibnamefont
  {Li}}, \bibinfo {author} {\bibfnamefont {Yufei}\ \bibnamefont {Ding}}, \ and\
  \bibinfo {author} {\bibfnamefont {Yuan}\ \bibnamefont {Xie}},\ }\bibfield
  {title} {\enquote {\bibinfo {title} {Tackling the qubit mapping problem for
  nisq-era quantum devices},}\ }in\ \href {\doibase 10.1145/3297858.3304023}
  {\emph {\bibinfo {booktitle} {Proceedings of the Twenty-Fourth International
  Conference on Architectural Support for Programming Languages and Operating
  Systems}}},\ \bibinfo {series and number} {ASPLOS '19}\ (\bibinfo
  {publisher} {Association for Computing Machinery},\ \bibinfo {address} {New
  York, NY, USA},\ \bibinfo {year} {2019})\ p.\ \bibinfo {pages}
  {1001–1014}\BibitemShut {NoStop}%
\bibitem [{\citenamefont {Spall}(1998)}]{SPSA}%
  \BibitemOpen
  \bibfield  {author} {\bibinfo {author} {\bibfnamefont {J.C.}\ \bibnamefont
  {Spall}},\ }\bibfield  {title} {\enquote {\bibinfo {title} {Implementation of
  the simultaneous perturbation algorithm for stochastic optimization},}\
  }\href {\doibase 10.1109/7.705889} {\bibfield  {journal} {\bibinfo  {journal}
  {IEEE Transactions on Aerospace and Electronic Systems}\ }\textbf {\bibinfo
  {volume} {34}},\ \bibinfo {pages} {817--823} (\bibinfo {year}
  {1998})}\BibitemShut {NoStop}%
\bibitem [{\citenamefont {Schollw\"ock}(2005)}]{DMRG}%
  \BibitemOpen
  \bibfield  {author} {\bibinfo {author} {\bibfnamefont {U.}~\bibnamefont
  {Schollw\"ock}},\ }\bibfield  {title} {\enquote {\bibinfo {title} {The
  density-matrix renormalization group},}\ }\href {\doibase
  10.1103/RevModPhys.77.259} {\bibfield  {journal} {\bibinfo  {journal} {Rev.
  Mod. Phys.}\ }\textbf {\bibinfo {volume} {77}},\ \bibinfo {pages} {259--315}
  (\bibinfo {year} {2005})}\BibitemShut {NoStop}%
\bibitem [{\citenamefont {Schuld}\ \emph {et~al.}(2019)\citenamefont {Schuld},
  \citenamefont {Bergholm}, \citenamefont {Gogolin}, \citenamefont {Izaac},\
  and\ \citenamefont {Killoran}}]{grad}%
  \BibitemOpen
  \bibfield  {author} {\bibinfo {author} {\bibfnamefont {Maria}\ \bibnamefont
  {Schuld}}, \bibinfo {author} {\bibfnamefont {Ville}\ \bibnamefont
  {Bergholm}}, \bibinfo {author} {\bibfnamefont {Christian}\ \bibnamefont
  {Gogolin}}, \bibinfo {author} {\bibfnamefont {Josh}\ \bibnamefont {Izaac}}, \
  and\ \bibinfo {author} {\bibfnamefont {Nathan}\ \bibnamefont {Killoran}},\
  }\bibfield  {title} {\enquote {\bibinfo {title} {Evaluating analytic
  gradients on quantum hardware},}\ }\href {\doibase
  10.1103/PhysRevA.99.032331} {\bibfield  {journal} {\bibinfo  {journal} {Phys.
  Rev. A}\ }\textbf {\bibinfo {volume} {99}},\ \bibinfo {pages} {032331}
  (\bibinfo {year} {2019})}\BibitemShut {NoStop}%
\bibitem [{\citenamefont {Takagi}\ \emph
  {et~al.}(2022{\natexlab{b}})\citenamefont {Takagi}, \citenamefont {Endo},
  \citenamefont {Minagawa},\ and\ \citenamefont {Gu}}]{ZEN_Takagi}%
  \BibitemOpen
  \bibfield  {author} {\bibinfo {author} {\bibfnamefont {Ryuji}\ \bibnamefont
  {Takagi}}, \bibinfo {author} {\bibfnamefont {Suguru}\ \bibnamefont {Endo}},
  \bibinfo {author} {\bibfnamefont {Shintaro}\ \bibnamefont {Minagawa}}, \ and\
  \bibinfo {author} {\bibfnamefont {Mile}\ \bibnamefont {Gu}},\ }\bibfield
  {title} {\enquote {\bibinfo {title} {Fundamental limits of quantum error
  mitigation},}\ }\href {\doibase 10.1038/s41534-022-00618-z} {\bibfield
  {journal} {\bibinfo  {journal} {npj Quantum Information}\ }\textbf {\bibinfo
  {volume} {8}} (\bibinfo {year} {2022}{\natexlab{b}}),\
  10.1038/s41534-022-00618-z}\BibitemShut {NoStop}%
\bibitem [{\citenamefont {Endo}\ \emph {et~al.}(2018)\citenamefont {Endo},
  \citenamefont {Benjamin},\ and\ \citenamefont {Li}}]{ZEN_Endo}%
  \BibitemOpen
  \bibfield  {author} {\bibinfo {author} {\bibfnamefont {Suguru}\ \bibnamefont
  {Endo}}, \bibinfo {author} {\bibfnamefont {Simon~C.}\ \bibnamefont
  {Benjamin}}, \ and\ \bibinfo {author} {\bibfnamefont {Ying}\ \bibnamefont
  {Li}},\ }\bibfield  {title} {\enquote {\bibinfo {title} {Practical quantum
  error mitigation for near-future applications},}\ }\href {\doibase
  10.1103/PhysRevX.8.031027} {\bibfield  {journal} {\bibinfo  {journal} {Phys.
  Rev. X}\ }\textbf {\bibinfo {volume} {8}},\ \bibinfo {pages} {031027}
  (\bibinfo {year} {2018})}\BibitemShut {NoStop}%
\bibitem [{\citenamefont {\emph{et al.}}(2021)}]{Qiskit}%
  \BibitemOpen
  \bibfield  {author} {\bibinfo {author} {\bibfnamefont {MD~SAJID~ANIS}\
  \bibnamefont {\emph{et al.}}},\ }\href {\doibase 10.5281/zenodo.2573505}
  {\enquote {\bibinfo {title} {Qiskit: An open-source framework for quantum
  computing},}\ } (\bibinfo {year} {2021})\BibitemShut {NoStop}%
\bibitem [{\citenamefont {Kraft}(1988{\natexlab{b}})}]{slsqp}%
  \BibitemOpen
  \bibfield  {author} {\bibinfo {author} {\bibfnamefont {Dieter}\ \bibnamefont
  {Kraft}},\ }\bibfield  {title} {\enquote {\bibinfo {title} {A software
  package for sequential quadratic programming, ein software-paket zur
  sequentiellen quadratischen optimierung, forschungsbericht. deutsche
  forschungs- und versuchsanstalt für luft- und raumfahrt, dfvlr},}\ }\href
  {https://www.tib.eu/de/suchen/id/TIBKAT%3A016896521} {\ \textbf {\bibinfo
  {volume} {88-28}} (\bibinfo {year} {1988}{\natexlab{b}})}\BibitemShut
  {NoStop}%
\bibitem [{\citenamefont {Gacon}\ \emph {et~al.}(2021)\citenamefont {Gacon},
  \citenamefont {Zoufal}, \citenamefont {Carleo},\ and\ \citenamefont
  {Woerner}}]{QNSPSA}%
  \BibitemOpen
  \bibfield  {author} {\bibinfo {author} {\bibfnamefont {Julien}\ \bibnamefont
  {Gacon}}, \bibinfo {author} {\bibfnamefont {Christa}\ \bibnamefont {Zoufal}},
  \bibinfo {author} {\bibfnamefont {Giuseppe}\ \bibnamefont {Carleo}}, \ and\
  \bibinfo {author} {\bibfnamefont {Stefan}\ \bibnamefont {Woerner}},\
  }\bibfield  {title} {\enquote {\bibinfo {title} {Simultaneous perturbation
  stochastic approximation of the quantum fisher information},}\ }\href@noop {}
  {\bibfield  {journal} {\bibinfo  {journal} {Quantum}\ }\textbf {\bibinfo
  {volume} {5}},\ \bibinfo {pages} {567} (\bibinfo {year} {2021})}\BibitemShut
  {NoStop}%
\end{thebibliography}
\end{document}